\documentclass[%
 aip,
 jcp,%
 amsmath,amssymb,
reprint,%
leqno, 
]{revtex4-2}

\usepackage{lineno} 

\usepackage{graphicx}
\usepackage{dcolumn}
\usepackage{bm}
\usepackage[super]{nth}
\usepackage[usenames,dvipsnames]{color}
\usepackage{full_name_references} 
\usepackage{flowchart} 
\usepackage{python_syntax}
\usepackage[noend]{algpseudocode}
\usepackage{algorithm}

\usepackage{ulem}
\begin{document}

\preprint{AIP/123-QED}

\title[Grid-based state space exploration]{Grid-based state space exploration for molecular binding}
\author{Hana Zupan}
\author{Frederick Heinz}
\author{Bettina G.~Keller}%
\email{bettina.keller@fu-berlin.de}
\affiliation{ 
Department of Biology, Chemistry, Pharmacy, Freie Universität Berlin, Arnimallee 22, 14195 Berlin, Germany
}

\date{\today}


%
%

\begin{abstract} 
Binding processes are difficult to sample with molecular-dynamics (MD) simulations. In particular, the state space exploration is often incomplete. Evaluating the molecular interaction energy on a grid circumvents this problem but is heavily limited by state space dimensionality. 
Here, we make the first steps towards a low-dimensional grid-based model of molecular binding.
We discretise the state space of relative positions and orientations of the two molecules under the rigid body assumption.
The corresponding program is published as the Python package \texttt{molgri}.
For the rotational component of the grids, we test algorithms based on Euler angles, polyhedra and quaternions, of which the polyhedra-based are the most uniform.
The program outputs a sequence of molecular structures that can be easily processed by standard MD programs to calculate grid point energies.
We demonstrate the grid-based approach on two molecular systems: a water dimer and a coiled-coil protein interacting with a chloride anion. 
For the second system we relax the rigid-body assumption and improve the accuracy of the grid point energies by an energy minimisation.
In both cases, oriented bonding patterns and energies confirm expectations from chemical intuition and MD simulations. 
We also demonstrate how analysis of energy contributions on a grid can be performed and demonstrate that electrostatically-driven association is sufficiently resolved by point-energy calculations. 
Overall, grid-based models of molecular binding are potentially a powerful complement to molecular sampling approaches, and we see the potential to expand the method to quantum chemistry and flexible docking applications.
\end{abstract}

\maketitle


\section{Introduction}
Molecular binding, i.e.~the process in which two or more molecules form a long-lived complex, is ubiquitous but very difficult to study, both experimentally and computationally. 
Binding processes range across several length and time scales. 
Examples include
dimers of small solvent molecules such as H$_2$O dimers or HF dimers \cite{buckingham1991hydrogen}, 
solvation shells around ions in solution \cite{rinne2014ion}, 
chelate-complexes that can mask more complex organic molecules \cite{reinmuth2019structural}, 
association of molecules on two-dimensional surfaces including biological membranes, 
binding of small molecular inhibitors to protein receptors \cite{plattner2015protein, casasnovas2017unbinding}, 
the formation of protein-protein complex \cite{kahler2020protein, wenz2021target}, and 
self-assembly of molecular monomers into almost macroscopic structure such as hydrogels or fibrils \cite{morriss2015computational, morozova2018electrostatic, hellmund2021functionalized}.
Using molecular-dynamics (MD) simulations combined with rare-event techniques \cite{bolhuis2002transition, barducci2011metadynamics,zuckerman2017weighted, bowman2010enhanced}, one can now model and analyze molecular binding at atomistic detail
\cite{bruce2018new, limongelli2020ligand, bruce2018new}.
However, MD simulations rely on sampling the conformational space and assume that the ergodic hypothesis is fulfilled. 
This is problematic for binding processes, because binding often occurs via multiple distinct binding pathways. 
These pathways predominantly manifest when the molecules are already in direct contact \cite{plattner2015protein, plattner2017complete, capelli2019exhaustive, lemke2018efficient}.
But even with large distances between associating molecules their long-range electrostatic interactions are not isotropic but can ``guide'' them towards forming a complex \cite{batra2013long, waldner2018electrostatic, kahler2020protein}. 
Exhaustively sampling all relative positions and relative orientations of the binding partners is challenging, and even determining whether or not the simulation has converged is difficult \cite{chodera2016simple}.  
An alternative to sampling is to discretize the conformational space on a grid and to explore its details by changing the resolution of the grid. 
Recently, methods that use this approach to calculate grid-based models of conformational dynamics have been proposed 
\cite{bicout1998electron, donati2018estimation, heida2018convergences, donati2021markov, donati2022review}.
Selecting the grid points so that interactions in all relevant degrees of freedom are resolved uniformly is a crucial step in this process.
We here make the first steps towards grids for the binding process of two molecules. 
As a first approximation, we model each of the molecules as a rigid body. 
Fixing the first molecule at the origin of the coordinate system, the relative positions and orientations of the second molecule can be described by a combination of three subspaces: 
($i$) translation in radial direction, 
($ii$) rotation around the origin, and 
($iii$) rotation around its centre of mass
(Fig.~\ref{fig:rotations}.A).
Discretising each of the three subspaces yields a regular grid which we call a trans-rot-rot grid.
Each of the grid-points can then be transformed into a set of matrices acting on position vectors of the atoms that yield a specific relative arrangement of the two molecules, i.e.~a molecular structure.
A critical point in this process is the construction of uniform rotational grids.
Since rotations can be represented as points on a unit sphere (Fig.~\ref{fig:rotations}.B), this problem is mathematically closely intertwined with uniformly distributing $N$ points on a sphere. 
We test rotation grids based on random and systematic Euler angles \cite{euler1776nova}, 
random quaternions \cite{kirk2012graphics}, polyhedra \cite{yershova2004deterministic, yershova2010generating} and polytopes \cite{karney2007quaternions}.

For the construction and analysis of rotational grids, we can build on a vast body of previous research \cite{diebel2006representing} since most areas of scientific computing need to consider grids on spheres or rotation spaces (e.g. geo\cite{purser2011standardized} and atmospheric \cite{sadourny1972conservative} sciences, robotics \cite{lindemann2004incremental}, computer animations \cite{kirk2012graphics}).
Rotation matrices and the corresponding quaternion representations of rotations are also frequently used in the context of molecular sampling, most notably for the superposition of molecular structures \cite{kneller1991superposition}. 
Other applications include
integrators of the equations of motion for rigid bodies \cite{fincham1992leapfrog, kol1997symplectic, miller2002symplectic},
reverse mapping in multiscale simulations \cite{nielsen2010recent},
analysis of rotational distributions \cite{stumpe2007aqueous}, rotational entropies \cite{heinz2019computing} or rotational correlation functions \cite{lynden1989reorientational}.
We implemented trans-rot-rot discretisation of association space in the program \texttt{molgri} and provide it as a free Python package. 
The program takes structure files of two molecules as inputs and returns the relative positions and orientations of the molecules on a trans-rot-rot grid.
The output is written in the format of a MD-trajectory, so that the grid point energies can be calculated using a standard MD program.
In this contribution, we briefly discuss the theory of rotations and explain the implementation and the usage of the program \texttt{molgri}.
Next, we test rotational grids generated by six different algorithms regarding their uniformity and suitability for molecular studies.
We then demonstrate how trans-rot-rot grids can be used to analyze a molecular binding process.
We discuss the position- and orientation-dependent contributions to the interaction energies of two water molecules, i.e.~molecules which are modelled as rigid bodies in classical force fields.
As a second example, we consider the interaction between a Cl$^{-}$-anion and a coiled-coil peptide \cite{hellmund2021functionalized}.

\section{Theory}
We introduce three common descriptions of rotations and briefly review how they relate to each other. We also discuss measures of uniformity applicable to rotation grids.
For a systematic overview of rotation representations and conversions between different formats we recommend a comprehensive overview article by Diebel \cite{diebel2006representing}.

\subsection{Rotation matrices}
A 3D rotation matrix $R$ is an orthogonal $3\times3$ matrix, i.e. a matrix which fulfills 
$R R^{\mathrm{T}} = R^{\mathrm{T}} R=I $,
where $R^{\mathrm{T}}$ is the transpose of $R$ and $I$ is the $3\times3$ identity matrix.
When applied to two vectors $\mathbf{v}'= R\mathbf{v}$ and $\mathbf{u}'= R\mathbf{u}$, an orthogonal matrix preserves the length of these vectors and the angle between them, which are two key properties of rotations.
The determinant of an orthogonal matrix always equals $\pm 1$. We define orthogonal matrices with $\det R = 1$ as those describing \textit{proper} rotations while those with $\det R = - 1$ describe \textit{improper} rotations or rotation-inversions. 

\subsection{Euler angles}
\label{sec:euler_angles}

Euler angles, first introduced by Leonhard Euler \cite{euler1776nova}, are based on the idea that ($i$) any two Cartesian axes in 3D space span a plane in which elementary rotations can be parametrized with a single angle, and ($ii$) any rotation in 3D space can be written as a sequence of three such elementary rotations around non-repeating axes.
%
%
%

The triple of angles $\mathbf{u}_{xyz} = [\phi, \theta, \psi]^T$ is called an Euler angle set and is a description of a 3D rotation in which a vector undergoes a sequence of three elementary (planar) rotations: first around axis $z$ for an angle $\psi$, then around $y$ for $\theta$ and finally around $x$ for $\phi$. Other sequences of axes are also possible, another common choice is the $zxz$ set.
%
%

%
Since the rotation axes for the elementary rotations are the axes of the Cartesian coordinate system, their corresponding elementary rotation matrices are simple trigonometric functions of the rotation angle. For rotation of $\phi$ around an x-axis we have:
\begin{equation}
    \label{eq:rot_matrices}
    R_{x}(\phi)= {\left[\begin{array}{ccc}1 & 0 & 0 \\ 0 & \cos (\phi) & \sin (\phi)\\ 0 & -\sin (\phi) & \cos (\phi)\end{array}\right]} ,
\end{equation}
and the $R_y$ and $R_z$ elementary matrices are simple permutations of the matrix above \cite{wiki_rot_matrix}.
To transform an Euler angle set into a single rotation matrix, we multiply the corresponding elementary rotation matrices in the correct order:
\begin{equation}
R_{x y z}(\phi, \theta, \psi):=R_{x}(\phi) R_{y}(\theta) R_{z}(\psi) .
\end{equation}

\subsection{Quaternions}
\label{sec:quaternions}
Another common representation of rotations is that of unit quaternions. Quaternions have been invented by William Rowan Hamilton \cite{hamilton1840new, hamilton1850quaternions} and are mostly known for their application in space rotations. A quaternion $\mathbf{q}$ is a 4-dimensional complex number

\begin{equation}
    \mathbf{q} = \left[q_{0}, q_{1}, q_{2}, q_{3}\right]^T = a+b \mathbf{i}+c \mathbf{j}+d \mathbf{k} ,
\end{equation}

where a, b, c, and d are real numbers and $\mathbf{i}$, $\mathbf{j}$, and $\mathbf{k}$ are the basic quaternions that can be interpreted as unit vectors along three perpendicular coordinate axes with the property $\mathbf{i}^{2}=\mathbf{j}^{2}=\mathbf{k}^{2}=-1$. The length of a quaternion is defined as

\begin{equation}
    \|\mathbf{q}\|=\sqrt{q_{0}^{2}+q_{1}^{2}+q_{2}^{2}+q_{3}^{2}}.
\end{equation}
A unit quaternion is a quaternion with $\|\mathbf{q}\| =1$.
We state, but omit the proof, that there always exist exactly two unit quaternions that map to one rotation matrix. 
For the derivation see Refs.~\onlinecite{hamilton1840new, hamilton1850quaternions, diebel2006representing}.
A rotation of any vector $\mathbf{v}$ to $\mathbf{v}^{\prime}$ can be described in terms of quaternion multiplication. 
For this, the three-dimensional vector $\mathbf{v}=[v_1, v_2, v_3]^T$ is interpreted as a quaternion in which the first element (the scalar element) is zero: $\mathbf{v} = 0 + v_1 \mathbf{i} + v_2 \mathbf{j} + v_3 \mathbf{k}$.
The rotation then is  
\begin{equation}
\label{eq:quat_mult}
    \mathbf{v}^{\prime}=\mathbf{q} \mathbf{v} \mathbf{q}^{-1} ,
\end{equation}
where $\mathbf{q}$ is a unit quaternion and 
\begin{equation}
\mathbf{q}^{-1}=\frac{\overline{\mathbf{q}}}{\|\mathbf{q}\|}, \qquad \qquad \overline{\mathbf{q}} = \left[ q_0, - q_1, -q_2, -q_3 \right]^T.
\end{equation}
is its reciprocal.
The multiplication of quaternions in eq.~\ref{eq:quat_mult} is defined by the Hamilton product \cite{hamilton1840new, hamilton1850quaternions, diebel2006representing}.

Quaternions have proven to be the most elegant solution in computational implementation of rotations. Since there is a natural and smooth mapping from the space of quaternions to the space of rotations (the $SO(3)$ group), interpolation of rotations can be easily performed in quaternion space and no singularities occur within this description, as opposed to the Euler angles parametrisation. 

\subsection{Rodrigues' formula}
Rotational matrices can be inferred from an initial vector $\mathbf{v}$ and rotated vector $\mathbf{v}'$ using Rodrigues' formula that has been independently discovered by Euler \cite{evlero1770problema} as well as Rodrigues \cite{rodrigues1840lois}. 
For the derivation, see chapter 3 of Ref.~\onlinecite{lynch2017modern}. 
In essence, we can calculate the unit vector $\boldsymbol{\hat{\omega}}$ along the axis of a rotation that transforms $\mathbf{v}$ into $\mathbf{v}'$ and the angle of rotation $\theta$ as
\begin{equation}
\label{eq:helper_rodrigues}
    \boldsymbol{\hat{\omega}} = \frac{\mathbf{v} \times \mathbf{v}'}{\|\mathbf{v}\|\|\mathbf{v}'\|},\quad
    \sin(\theta) = \|\boldsymbol{\hat\omega}\|,\quad
    \cos(\theta) =\frac{\mathbf{v} \cdot \mathbf{v}'}{\|\mathbf{v}\|\|\mathbf{v}'\|}.
\end{equation}
We additionally define the skew-symmetric matrix of a vector $
\boldsymbol{\omega} =\left[\begin{array}{lll}
\omega_{1} &\omega_{2} & \omega_{3}
\end{array}\right]^{\mathrm{T}} \in \mathbb{R}^{3}
$ as
\begin{equation}
\label{eq:skew}
[\omega]_{\times} =\left[\begin{array}{ccc}
0 & -\omega_{3} & \omega_{2} \\
\omega_{3} & 0 & -\omega_{1} \\
-\omega_{2} & \omega_{1} & 0
\end{array}\right].
\end{equation}
The transformation $\mathbf{v} \rightarrow \mathbf{v}'$ is then performed with
\begin{equation}
\label{eq:rodrigues}
\mathbf{v} '=e^{[\hat{\omega}]_{\times} \theta}\mathbf{v} =(I+\sin \theta[\hat{\omega}]_{\times}+(1-\cos \theta)[\hat{\omega}]_{\times}^{2} )\mathbf{v},
\end{equation}
which is known as Rodrigues' rotation formula. Writing the skew-symmetric matrices in eq.~\ref{eq:rodrigues} in matrix form, we can directly convert this representation into a rotation matrix multiplication.

\subsection{Grid coverage and uniformity}
\label{sec:theory_uniformity}

We use different parameterisations of rotations to create rotation grids. In order to compare how well the generated set of grid points covers the entire space of rotations, we introduce a measure of uniformity. 
Consider a rotation grid with $N$ grid points, each represented by a point on a unit sphere (radius $r=1$, total surface area $A = 4 \pi $). We calculate the local grid density in various sectors of the unit sphere by counting the number of grid points $N(S(\mathbf{v}, \alpha))$ within a well-defined sub-area $S(\mathbf{v}, \alpha)$ of the sphere surface and calculating the coverage ratio
\begin{eqnarray}
    c(\mathbf{v}, \alpha) &=& \frac{N(S(\mathbf{v}, \alpha))}{N}.
\end{eqnarray}
The vector $\mathbf{v}$ denotes the center of $S$, and the parameter $\alpha$ determines its surface area $A_S$.
In a fully uniform grid, the coverage would be equal for all sub-areas of equal size $A_S$ and further equal to the ratio $A_S/A$.

\begin{figure}[h!]
\includegraphics[width=7cm]{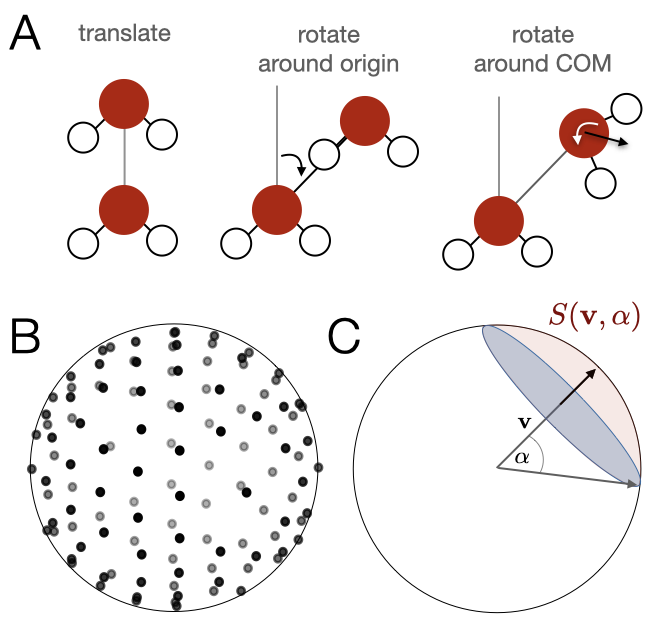}
\centering
\caption[Spherical cap]{\textbf{A.} Three types of motion in molecular binding: radial translation, relative rotation and internal rotation of one molecule around its own centre of mass. \textbf{B.} Example of a rotation grid with $N_{rot}$ grid points. \textbf{C.} Measuring local grid density. Spherical cap with area $S(\mathbf{v}, \alpha)$ (shaded) is defined by an axis vector $\textbf{v}$ and an angle $\alpha$. The ratio of grid points within spherical cap $N(S(\mathbf{v}, \alpha))$ with $N_{rot}$ is defined to be local grid density.}
\label{fig:rotations}
\end{figure}
A suitable choice of sub-areas is a class of curved surfaces called spherical caps\cite{wiki_spherical_cap}, defined by the intersection of the unit sphere with a plane (\figref{rotations}C).
The intersection itself is a circle, called the base of the spherical cap.
Here, $\mathbf{v}$ is the unit vector that connects the center of the unit sphere with the center of the cap and defines the center of $S$.
The parameter $\alpha$ is determined by the opening angle of the spherical cap, i.e.~the angle $\alpha$ between $\mathbf{v}$ and a unit vector that points to rim of the base, and is the quantity defining the size of $S$. 
The area of a spherical cap is \cite{wiki_spherical_cap}
\begin{equation}
\label{eq:spherical_cap}
   A_S = 2 \pi (1-\cos\alpha).  
\end{equation}
And so the expected coverage for a uniform grid is
\begin{eqnarray}
\label{eq:ideal_ratio}
    c_{\mathrm{uniform}}(\alpha) &=& \frac{A_S}{A} =  \frac{1}{2}(1-\cos\alpha)
\end{eqnarray}
See Refs.~\onlinecite{yershova2010generating, yershova2004deterministic} for a more detailed discussion on measures of uniformity.

\section{Software}
\label{sec:software}
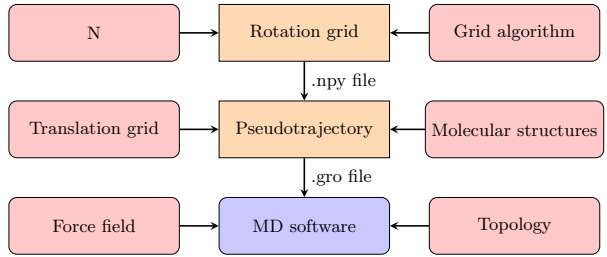
\begin{figure}[h!]
\centering
\resizebox{8cm}{!}{%
\begin{tikzpicture}[node distance=1.7cm]

\node (pro1) [process] {Rotation grid};
\node (in1) [startstop, right of=pro1, xshift=2cm] {Grid algorithm};
\node (start) [startstop, left of=pro1, xshift=-2cm] {N}; 
\node (pro3) [process, below of=pro1] {Pseudotrajectory};
\node (pro2a) [startstop, left of=pro3, xshift=-2cm] {Translation grid};
\node (pro3b) [startstop, right of=pro3, xshift=2cm] {Molecular structures};
\node (pro4) [result, below of=pro3] {MD software};
\node (pro4a) [startstop, left of=pro4, xshift=-2cm] {Force field};
\node (pro2b) [startstop, right of=pro4, xshift=2cm] {Topology};
\draw [arrow] (pro4a) -- (pro4);
\draw [arrow] (pro2a) -- (pro3);
\draw [arrow] (pro3b) -- (pro3);
\draw [arrow] (start) -- (pro1);
\draw [arrow] (in1) -- (pro1);
\draw [arrow] (pro1) -- (pro3)  node[midway,right] {.npy file};
\draw [arrow] (pro3) -- (pro4)  node[midway,right] {.gro file};
\draw [arrow] (pro2b) -- (pro4);
\end{tikzpicture}
}
\caption[Project workflow]{Workflow of grid-based state space exploration. The first two stages are performed with the \texttt{molgri} package, the resulting pseudo-trajectories can be further evaluated by most MD programs.} 
\label{fig:workflow}
\end{figure}

The Python package  \texttt{molgri} is freely available within the framework of Python Package Index \cite{molgri_pypi}.
Documentation and source code are available via GitHub \cite{molgri_GitHub}.
Dependencies are listed in section I of the supplementary information.
The package \texttt{molgri} implements a range of generation, analysis and plotting tools. 
A short user guide is provided in \secref{using_package}. 
As the workflow in \figref{workflow} indicates, the main output of \texttt{molgri} are the of relative positions and orientations of two molecules on a trans-rot-rot grid. 
The output is written in the file-format of a GROMACS molecular-dynamics trajectory (\texttt{.gro}-file), such that is can be further processed by standard MD programs to obtain and analyse point energies.
In this contribution, we use GROMACS for the energy calculation, but since the \texttt{.gro}-file format is a standard file-format for MD trajectories and can readily be converted in a other formats, the user is free to use the MD program of their choice for this step.
We use the term pseudo-trajectory for the output, because it has the file format of a MD trajectory (\texttt{.gro} file), but it represents a trans-rot-rot grid and not the result of MD sampling.
\texttt{molgri} generates pseudo-trajectories in two steps.
In the first step, the program generates a rotation grid, that is, a collection of $N_{\mathrm{rot}}$ points distributed on a unit sphere. 
We implemented and tested six algorithms for this step (see section \ref{sec:grid_alg}).
In the second step, the rotation grid is combined with a translation grid with $N_{\mathrm{trans}}$ elements. 
Overall we obtain $N_{\mathrm{trans}}\cdot N_{\mathrm{rot}} \cdot N_{\mathrm{rot}}$ relative positions and orientations for the two-molecule system (see section \ref{sec:grid_to_traj}).

We took particular care to build a modular and flexible program structure. Therefore, each step outputs a standard file format that can be reused by other applications. This also means that a rotation grid of particular size can be reused for any pair of molecules.

\subsection{Step 1: rotational grid algorithms \label{sec:grid_alg}}
The six algorithms we implemented to generate rotational grids are: 
Systematic Euler grid, 
Random Euler angles, 
Random unit quaternions \cite{kirk2012graphics}, 
3D cube grid \cite{yershova2004deterministic}, 
Icosahedron grid \cite{yershova2010generating}, and 
4D cube grid \cite{karney2007quaternions}. 
The algorithms are detailed in section II of the supplementary information, but we sketch the ideas behind them here.
Systematic Euler grid is based on a uniformly spaced grid between 0 and $2\pi$ for each of the three Euler angles using the $xyz$ sequence of axes.
Random Euler angles algorithm generates a random rotation by selecting a uniform random number in the $[0, 2\pi]$ interval for each of the three Euler angles.
Random unit quaternions algorithm generates random rotations by selecting random quaternions using the subgroup algorithm \cite{kirk2012graphics}.
Icosahedron grid and 3D cube grid use as a regular grid on the surface of a polyhedron to approximate a uniform grid on a unit sphere. 
This approach starts with inscribing a 3D polyhedron (a cube\cite{yershova2004deterministic} or an icosahedron\cite{yershova2010generating}) into a unit sphere, then constructing a grid on the faces of the polyhedron and projecting the grid points onto the unit sphere. 
The number of grid points $N_{\mathrm{rot}}$ can be increased by subdividing the faces of the polyhedron.
We subdivide square faces (cube) on a layered diagonal square lattice and triangle faces (icosahedron) on a layered triangle lattice. 
4D cube grid\cite{karney2007quaternions} extends this concept to polytopes (generalization of polyhedra to higher dimensional spaces) and quaternions.
In this case, a four-dimensional hyper-cube is inscribed into a four-dimensional unit hyper-sphere, a grid is constructed on its faces, and the grid points are projected onto the hyper-sphere.
The resulting four-dimensional vectors are interpreted as quaternions.
Some of the six algorithms generate a set of Euler angles, others a set of quaternions and still others a set of points on a sphere. We unify all outputs in the latter format by letting the quaternions or Euler angles act on a unit vector in the $z$-direction and saving the rotated vector as a three-dimensional grid point (algorithm I in section II of the supplementary information).
Systematic Euler grid and polyhedron/polytope-based grid algorithms use regular grids to generate the rotations, and can therefore natively only generate rotational grids with specific numbers of grid points $N_{\mathrm{rot}}$.
To provide complete flexibility in the number of grid points $N_{\mathrm{rot}}$, we first generate a grid of size $M \geq N_{\mathrm{rot}}$, then order all points in a way that maximizes coverage and truncate at $N_{\mathrm{rot}}$ (algorithm II in section II of the supplementary information).

\subsection{Step 2: from rotation grids to pseudo-trajectories}
\label{sec:grid_to_traj}

In the second step, the translational grid, the rotational grid of relative positions and the rotational grid of relative orientations are combined. 
We use a linear equidistant grid for the translations, and currently the same rotational grid for selection of relative positions and relative orientations.
Simplifications due to symmetry are not taken into consideration.
A full grid in state space is a product (meshgrid) of all three component grids.
To generate a pseudo-trajectory, the rotational grid is first transformed into a series of rotation matrices using Rodrigues formula (\equref{rodrigues}).
Specifically, the unit-vector in the $z$-direction is the initial vector $\mathbf{v}$ and the grid point is the rotated vector $\mathbf{v}'$.
Rotation axis and angle can then be calculated using \equref{helper_rodrigues}.
The pseudo-trajectory is generated by keeping molecule 1 fixed at the origin of the three-dimensional coordinate system while molecule 2 is translated and rotated.
Initially, the two molecules are positioned in their starting orientations at distance $r_0$ (first value of the translational grid). 
The $N_{\mathrm{rot}}$ rotation matrices are applied to the second molecule as rotations around the origin and the resulting configurations are recorded. 
In each of the configurations, the $N_{\mathrm{rot}}$ rotation matrices are then applied to the second molecule as rotations around the center of mass, generating $N_{\mathrm{rot}}$ relative orientations. 
The process is repeated for each radius provided by the translational grid. 
All resulting configurations are recorded in a single \texttt{.gro} file.

\subsection{Using the \texttt{molgri} package}
\label{sec:using_package}
In this section, we provide short instructions for the installation of \texttt{molgri} package and its use from command line.
%
%
Assuming a compatible Python installation is already present, the package is installed with the command 
\begin{lstlisting}[style=DOS]
pip install molgri
\end{lstlisting}
After installation, \texttt{molgri} can be imported as a package into a Python program as \texttt{import molgri}. 
Alternatively, we also provide three scripts that can be run directly from the command line: 1) \texttt{molgri-io} that creates a standard tree of input/output directories, 2) \texttt{molgri-grid} that generates rotation grids including their plots, animations and statistical analyses and 3) \texttt{molgri-pt} that generates pseudotrajectory files. 
All three scripts can be run with an optional flag \texttt{--help} that returns a short user guide to the script.
Running the script \texttt{molgri-io} with the optional flag \texttt{--examples} 
\begin{lstlisting}[style=DOS]
molgri-io --examples
\end{lstlisting}
provides example input .gro files including all molecules and ions used in Results section of this contribution and generates the following directory structure 
\begin{lstlisting}[style=DOS]
.
 |-output
 | |-grid_files
 | |-pt_files
 | |-figures
 | |-animations
 | |-statistics_files
 |-input
 | |-CL.gro
 | |-H2O.gro
 | |-NA.gro
 | |-NH3.gro
 | |-example_protein.gro
\end{lstlisting}
Instead of running this script, the user can also manually create an \texttt{input/} directory and copy any \texttt{.gro} files for which pseudo-trajectories should be generated there. 

The second script \texttt{molgri-grid} implements step 1 from \figref{workflow}.  It is necessary to specify the number of grid points \texttt{-N} and the algorithm 
\texttt{-algo} (options: systemE, randomE, randomQ, cube4D, cube3D, ico) to generate a rotation grid. 
In addition, optional flags to this script are: \texttt{--draw} that saves a plot of grid points, \texttt{--animation} for the corresponding 3D animation, \texttt{--statistics} that generates a number of files and figures analysing the uniformity and convergence of this grid, and \texttt{--readable} that saves the grid points in a human-readable .txt format in addition to the standard .npy format.
The generated files are saved to the correspondingly named subfolder within the output folder.
For example,
\begin{lstlisting}[style=DOS]
molgri-grid -N 250 -algo ico --draw 
--animate --statistics --readable
\end{lstlisting}
generates a rotational grid with 250 grid points using the algorithm Icosahedron grid, draws and animates the grid, calculates the grid statistics, and additionally stores the grid in human-readable format.
The last command line script \texttt{molgri-pt} implements step 2 and creates a pseudotrajectory. 
It expects two molecular structure files in the \texttt{.gro}-format which should be stored in the \texttt{input/} folder. 
Usually it is preferential to select the smaller molecule as the \texttt{-m2} option since this is the molecule undergoing translations and rotations.
In addition, the user needs to specify the rotational grid in the form \texttt{-rot algorithm\_N} (see algorithm names above) and the translational grid (flag \texttt{-trans}).
The translational grid can be specified in one of the following formats: a list of distances, linspace(start, stop, num) or range(start, stop, step), where the units are in nanometers.
For example, the translational grid with grid points 1.0, 1.5, 2.0, 2.5, and 3.0 nm can be specified as
\begin{itemize}
    \item \texttt{-trans "(1, 1.5, 2, 2.5, 3)"}
    \item \texttt{-trans "linspace(1, 3, 5)"}
    \item \texttt{-trans "range(1, 3.1, 0.5)"}
\end{itemize}

In summary, the command
\begin{lstlisting}[style=DOS]
molgri-pt -m1 NH3 -m2 H2O -rot cube3D_15
-trans "linspace(1, 3, 5)"
\end{lstlisting}
generates a pseudo-trajectory of a water molecule rotating around an ammonium molecule, where the rotational grid is generated with a cube 3D algorithm and contains 15 points (or is read from /output/grid\_files/cube3D\_15.npy if already generated). The two structure files are expected as
./input/H2O.gro and ./input/NH3.gro, and the translational grid is the one discussed above. Finally, the user may select an optional flag \texttt{--only\_origin} which suppresses the generation of different orientations of the second molecule and only returns $N_{\mathrm{rot}}\cdot N_{\mathrm{trans}}$ structures. This is useful if the second molecule is spherically symmetric (e.g. a single ion or an atom).
%

%

\section{Results}

\subsection{Comparison of rotational grid algorithms}
%

    \begin{figure*}[t]
        \centering
        \includegraphics[width=\linewidth]{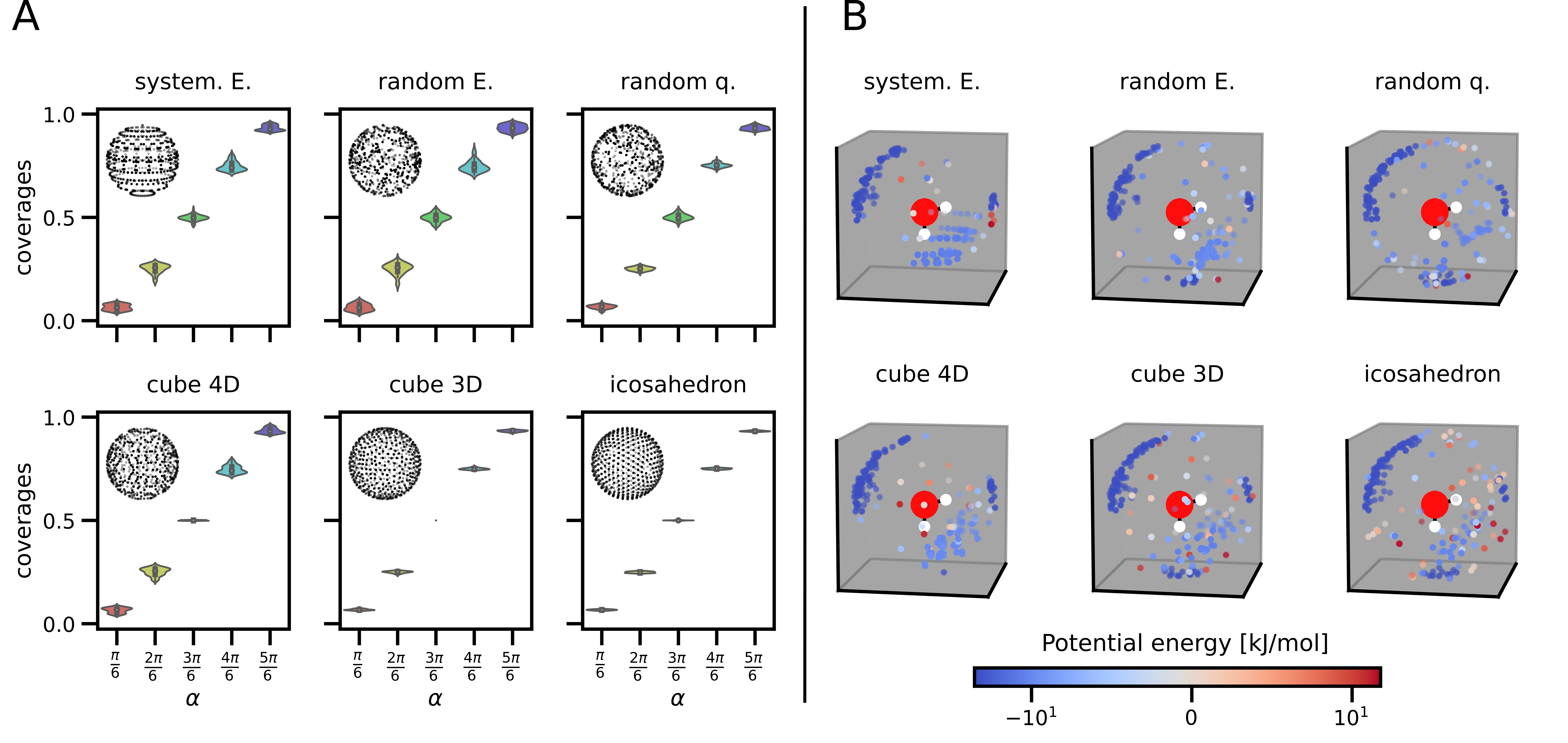}
        \caption{Comparison of rotation grid algorithms. \textbf{A.} Rotation grids with N=600 points are shown in upper left corners. The variance in grid densities across sphere surface is displayed as a set of violin plots. As a measure of uniformity we select 100 random axes and measure the density of the grid in this region by counting the number of points $N_{sc}$ that fall within an angle $\alpha$ of the axis and calculating the coverage $N_{sc}/N$. Broad distributions suggest that coverage is not uniform. \textbf{B.} The same rotation grids are used to create pseudo-trajectories for the water-water system. Potential energies [kJ/mol] of structures with various orientations and positions of the second water molecule are plotted, energy minima shown in dark blue. The Systematic Euler algorithm misses one of the three minima (below the hydrogen) and the cube 4D algorithm covers it with only one point, demonstrating the errors that may arise when using grids with poor uniformity.}
        \label{fig:violin_alpha}    
    \end{figure*}   

%
Algorithms presented in \secref{grid_alg} provide several ways to discretise rotation space, each generating a particular set of $N_{\mathrm{rot}}$ points on a sphere. 
The insets in \figref{violin_alpha}~A show the grid points generated by each of the algorithms for $N_{\mathrm{rot}}=600$. 
One can easily distinguish grid-based approaches from stochastic ones by the regularity and the symmetry of the grids. 
Note that in the Systematic Euler grid, the grid points align in circles along the ``latitudes'' around the sphere and are denser at the ``equator'' than at the ``poles''.
Since the random Euler angle algorithm is a direct stochastic analogue of the Systematic Euler grid, it also over-samples along the equator and under-samples the pole areas. 
Random unit quaternions algorithm samples from uniform distribution in the unit sphere (analytical proof in \cite{kirk2012graphics}), but due to fluctuations in a random sample with finite number of points, this does not necessarily yield the most uniform grid.
Polyhedra-based algorithms (3D cube grid and Icosahedron grid) show obvious regularities in their structure.
To quantitatively compare the grids, we recall the measure of uniformity introduced in section \secref{theory_uniformity} where the local density of grid points was measured by counting the number of grid points within a spherical cap area (\equref{spherical_cap}) and evaluating variations from ideal grid density given in \equref{ideal_ratio}. Local density is measured around 100 randomly chosen axes and the variation of measurements shown in form of a violin plot in \figref{violin_alpha}~A. The parameter $\alpha$ indicates the size of the spherical cap area in which counting is performed.
%
%
The narrower the distribution, the more uniform the grid.
The two algorithms that are based on inscribing polyhedra into three-dimensional sphere, 3D cube grid and Icosahedron grid, produce by far the most uniform grids across all values of values of $\alpha$.

It is important to consider several different angles to test uniformity at all scales, from narrow areas surrounding the axis to ones covering almost entire sphere (we test angles ranging from $\pi/6$ to $5\pi/6$). How variance changes with $\alpha$ provides additional information. We observe, for example, that variance in random Euler angle grid changes with $\alpha$, suggesting that the algorithm fails to achieve true randomness (random quaternion algorithm performs better in this regard). Using a single alpha value can also be deceiving as it may reflect inherent symmetries in grid construction. Notable examples are cubic grids with essentially nonexistent variance in grid densities for $\alpha=\pi/2$. For the 3D and 4D cubic grids, the subdivision is performed in the same way for each of the eight faces, meaning that each spherical cap covering a quarter of the sphere ($\alpha=\pi/2$) always contains the same collection of points ($\pm 1$ per face for $N$ not divisible by 8).
%

Grid uniformity is not only of theoretical interest but has real consequences when using rotation grids to discretise the space of molecular rotations. \figref{violin_alpha}~B displays a simple molecular example where the grids discussed above were employed to study a system of two water molecules. In this figure, areas where structures with low potential energy were found are shown in dark blue color. In most cases, we observe three minima: a broad one near the free electron pairs of the central oxygen atom and two more pointed ones near each of the central hydrogens, indicating three possible hydrogen bonding patterns. The difference between grid algorithms becomes apparent when we note that one of the three minima (near bottom hydrogen) is completely missing in the Systematic Euler grid and is only represented by a single point in the 4D cube grid. Although a very dense (600 points) grid was used, grids with poor uniformity failed to identify all regions of interest even in this simple example. This underlines the necessity of examining topology when selecting parametrisation of non-Euclidean spaces.


\begin{figure}[h!]
    \centering
    \includegraphics[width=8cm]{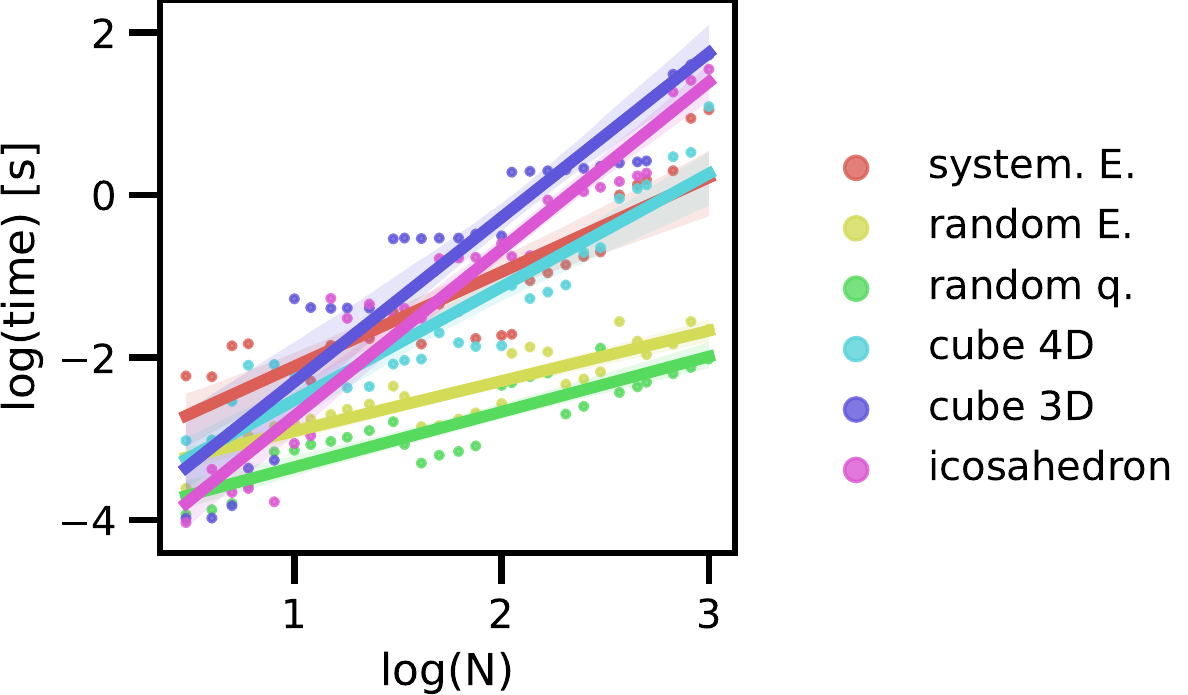}
    \caption[Time to construct rotation grids]{Time needed to construct rotation grids using six different algorithms. Polyhedra grids (icosahedron, cube 3D) are orders of magnitude more time intensive than random grids (random Euler angles, quaternions), but total time needed to create grids with up to 1000 points is still under 100 seconds.}
    \label{fig:time_algo}
\end{figure}

\figref{time_algo} compares how time-intensive the six rotational grid algorithms are. 
The timings have been measured on a Debian GNU/Linux 11 operating system, equipped with an Intel(R) Core(TM) i5-8500 CPU @ 3.00GHz and 30 GB RAM. 
Both random algorithms, Random Euler angles and Random unit quaternions, are very efficient, while the 3D cube grid and Icosahedron grid show much poorer scaling with the number of grid points.
The almost constant timing for long intervals of $N_{\mathrm{rot}}$ originate from the fact that polyhedra-based algorithms do not natively yield grids with arbitrary number of grid points - larger grids are created and truncated according to an unification algorithm found in Section II of the Supplementary information. 
However, even for the more time-consuming algorithms the computational costs are in the range of seconds for grids with up to 1000 points, which is very small compared to the cost of the entire analysis, which is dominated by the energy calculation along the pseudo-trajectory.
In summary, polyhedra-based grids are remarkably uniform. 
The somewhat larger computational cost for generating these grids are well worth the improved quality.

\subsection{Water-water system}
In the second step of the \texttt{molgri} procedure, rotation and translation grids are combined to systematically parameterise all three movements with which we parameterise association space - translation, relative rotation and internal rotation, see \figref{rotations}~A. After selecting two molecules, a pseudo-trajectory based on the meshgrid of the three grids is generated.

We first demonstrate the use of pseudo-trajectories on a system of two water molecules, since this is a small, well-studied example that nonetheless displays several oriented hydrogen-bonding patterns, enabling us to study how radial and rotational degrees of freedom contribute to potential energy. Moreover, rigid models of water are common in classical MD, meaning that a direct comparison with established methods is possible.




\begin{table}[h]
\begin{tabular}{lcc}
\hline & \text { Potential }[$\mathrm{kJ} / \mathrm{mol}$] &  \text {H-bond length [nm] } \\
\hline \text { systematic Euler a. } & --27.67 &  0.28 \\
\text { icosahedron grid } & --27.39   & 0.28 \\
\text { random Euler a. } & --27.35 &  0.28 \\
\text { random quaternions } & --27.24 &  0.27 \\
\text { 3D cube grid } & --27.23 &  0.27 \\
\text { \textbf{MD run} } & \textbf{--27.09}  & \textbf{0.28} \\
\text { 4D cube grid } & --27.03 & 0.27 \\
\hline
\end{tabular}
    \caption{Comparing minima of (pseudo)trajectories, each of the methods evaluating 60.000 points in state space.}
    \label{tab:water_min_energies}
\end{table}

In \tabref{water_min_energies} we perform such a comparison of minimal potential energies found in a relatively sparse ($N_{rot}=100$, $N_{trans}=5$, 0.3-0.32 nm) pseudo-trajectory of two rigid (TIP3P) water molecules with a standard GROMACS trajectory with the same number of frames. Since all found minima featured a hydrogen bond, we also compare hydrogen bond lengths. While there is some difference between different rotation algorithms, all minima fall within 2.5\% of the comparison calculation and all but one \texttt{molgri} algorithms even find a deeper minimum than a simulation with the same number of points. The plots in previous section (\figref{violin_alpha}) also suggest that the minima found by \texttt{molgri} occur at locations expected by partial charge distribution in water molecules. 




\begin{figure}[h]
    \centering
    \includegraphics[width=8cm]{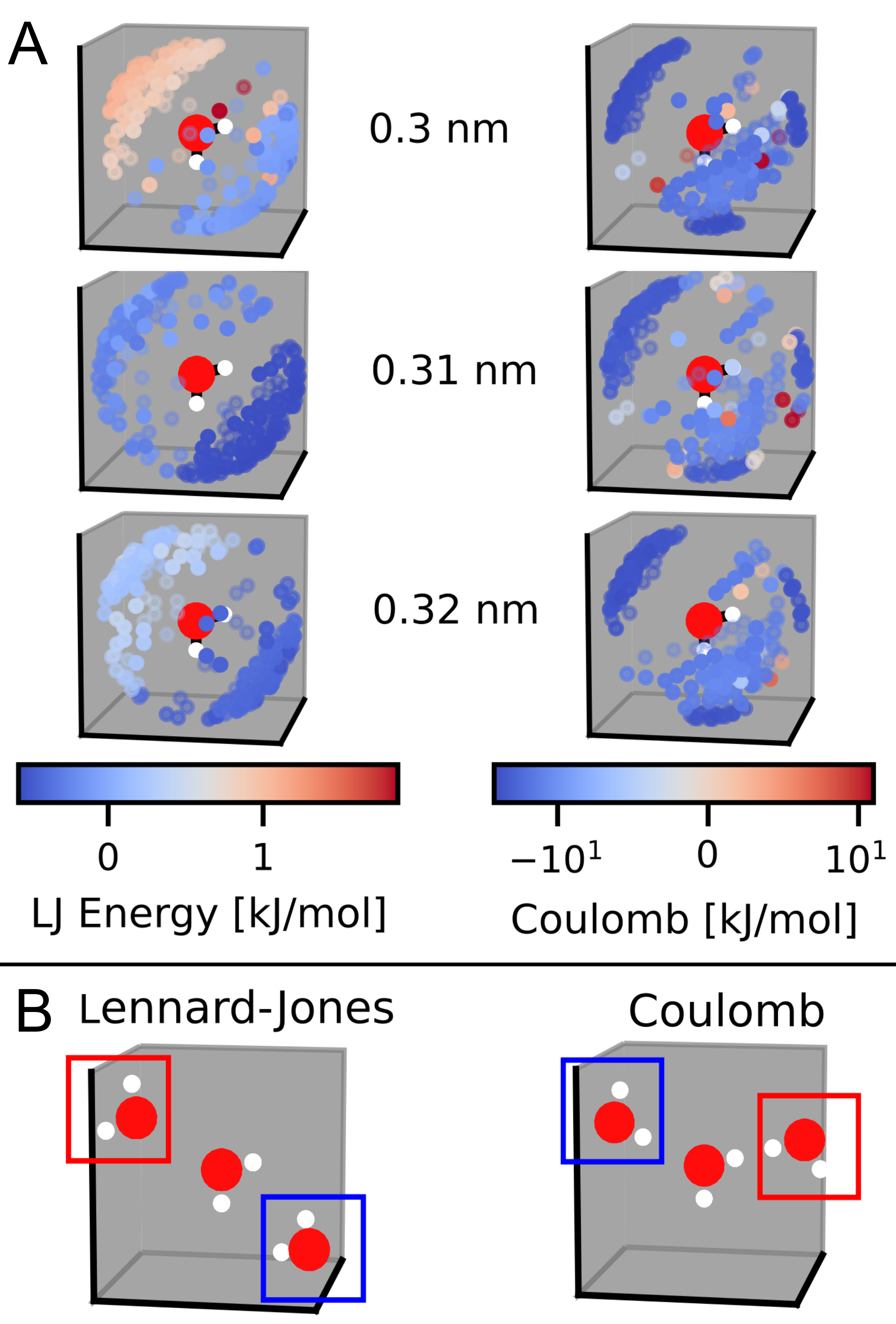}
    \caption{\textbf{A.} Coulomb and LJ energies [kJ/mol] for the full pseudotrajectory (icosahedron grid, $N_{\mathrm{rot}}=600$, $N_{\mathrm{trans}}=5$) separated into three plots based on the the radial COM-COM distance of both water molecules. \textbf{B.} Pseudo-trajectory structures with lowest (blue box) and highest (red box) Coulomb/LJ contributions.}
    \label{fig:radii_coulomb}
\end{figure}

In the the next step, we examine how different components of potential energy, specifically the Coulomb and Lennard-Jones contributions, vary with distance and orientation. The ability to systematically examine spatial dependence of energy contributions is a unique feature of our approach since classical trajectories rarely enable us to examine the full range of states, strongly biasing the sample towards a few easily accessible minima. For this examination, we construct longer pseudo-trajectories of the same system using Icosahedron grid with $N_{rot}=600$ and $N_{trans}=5$ between 0.3 and 0.32 nm, generating $600 \cdot 600 \cdot 5 = 1.8$ million time steps in total. 

In order to separate translational variation from rotational, we plot three shells with different radial distances in separate subplots of \figref{radii_coulomb}~A. We plot a point at the center of mass (COM) of the second water molecule and color-code it according to its Coulomb or Lennard-Jones energy. Since only a subset of the 1.8 million structures can be reasonably visualised, we first perform two levels of selection: 1) among all structures with the same COM but different internal orientations, we select the one with lowest energy and 2) of the points selected in step 1), 100 lowest and 100 highest energies are plotted. On the bottom of the same Figure, \ref{fig:radii_coulomb}~B, we plot the structures with the lowest and highest energy contribution so that internal orientation of the second water molecule can also be seen.

Completely different patterns arise for the two energy components. In the case of Lennard-Jones potential, radial distance plays a major role (as can be expected from the functional form of this potential) while the hydrogen-bonding-specific patterns are not present. Hydrogen-hydrogen interaction is the most favourable LJ interaction in this system while oxygen-oxygen one, especially at smaller radii, is most repulsive.

On the other hand, Coulomb energies barely vary with radial distance but a strong and persisting orientational pattern is observed instead. This could again be predicted by the form of Coulomb potential which scales with $1/r$, $r$ being the distance between charges. The lowest- and highest energy structures confirm that the cause is electrostatic repel between positively charged hydrogens and attractive interaction of a hydrogen with free electron pairs on an oxygen.

While the contribution analysis presented here is anything but novel for this simple system, we expect this kind of systematic study to be valuable for complex systems featuring a mix of interaction potentials, something that is very typical for protein systems.

\subsection{Polypeptide-ion system \label{sec:protein}}
%
        \begin{figure*}[t]
        \includegraphics[width=\textwidth]{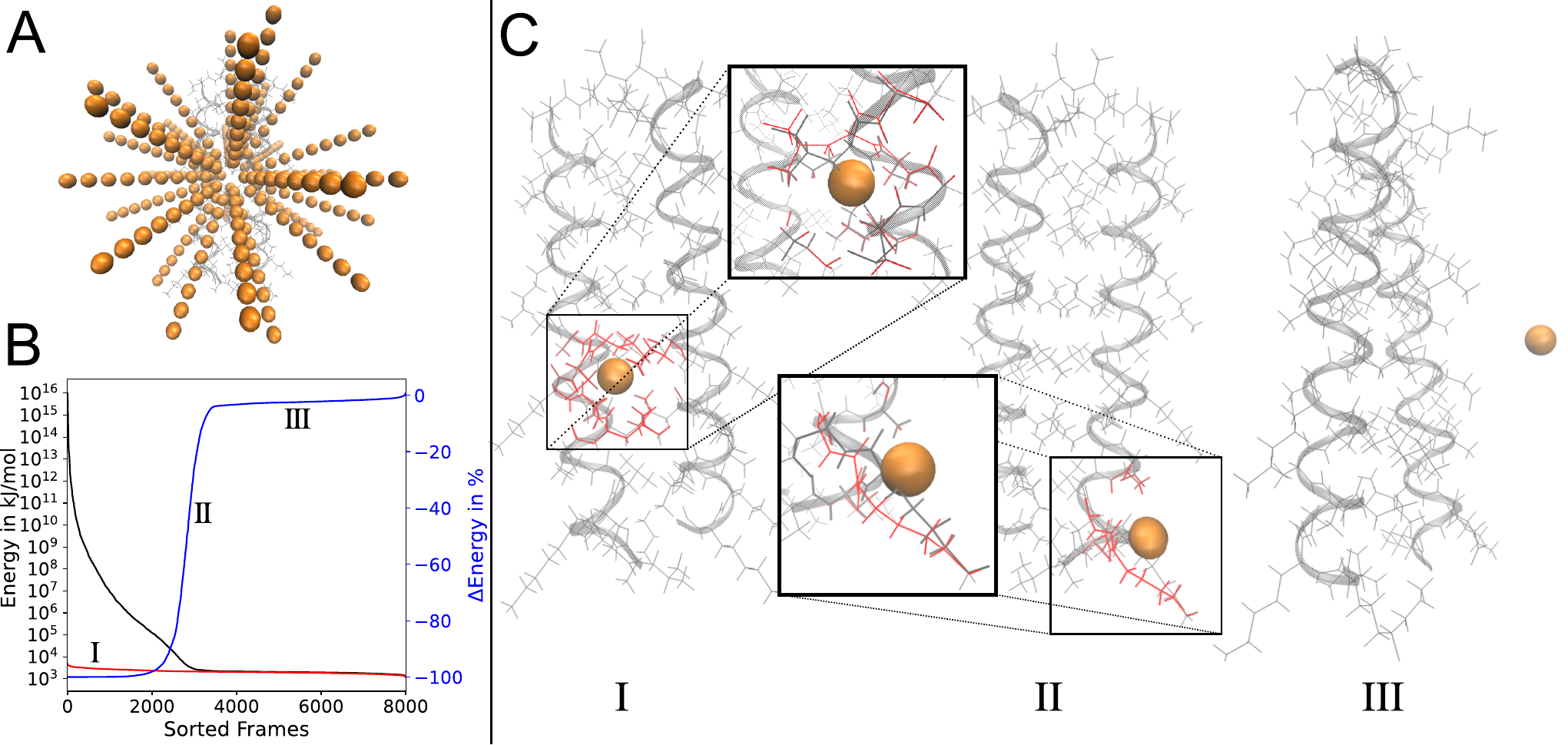}
        \caption{\textbf{A.} An example of a pseudo-trajectory with a sparse grid ($N_{\mathrm{rot}}=30$, $N_{\mathrm{trans}}=8$, distances 0.5 - 2.25 nm). Molecular structure of the coiled-coil peptide is shown in grey and the probe Cl$^-$ ion in orange. All frames of the pseudo-trajectory are shown overlapped, each contains one Cl$^-$ ion. In B and C parts of the figure, $N_{\mathrm{rot}}=1000$ is used, otherwise, the set-up is equivalent. \textbf{B.} Change in the potential energy of the coiled-coil-Cl$^-$ system through the usage of energy minimisation of each frame. Potential energy before (black) and after (red) energy minimization is shown for frames ordered from highest to lowest energy. Additionally, the relative change in energy by minimization is shown in blue. Some structures (zone I) exhibit immense relaxation during the minimisation; in these cases the initial structure featured overlapping Lenard-Jones potentials. Other structures (zone III) remain essentially unchanged.
        \textbf{C.} Example pseudo-trajectory frames from zones I, II and III. The initial pseudo-trajectory structure is shown in grey and the structure after minimisation in red. Only the movement of atoms within 0.6 nm of the Cl$^-$ ion are visualized in red; changes outside this sphere were minimal fluctuations.}
        \label{fig:Closeup}
        \end{figure*}    

%
The water-water dimer is an often modeled system without any internal degrees of freedom and can be treated accurately as a system of two rigid bodies. 
To test the implications of the rigid-body assumption on a system with many internal degrees of freedom, we use \texttt{molecularRotationalGrids} to evaluate the system of a chlorine anion interacting with a coiled-coil peptide (\figref{Closeup}).
The coiled-coil peptide hFF03 \cite{hellmund2021functionalized}  is a parallel homo-dimer of two peptide strands, each of which forms an $\alpha$-helix.
The resulting coiled-coil dimer is stabilized via a leucine zipper \cite{woolfson2005design}.
The ends of the peptide strand are not capped and the peptide is represented at pH=~7.
Consequently, the amid group of the N-terminal end of the coiled-coil dimer and amine group of the lysine side chains are positively charged.
Whereas the carboxyl group of the C-terminal end and the glutamatic acid side chains are negatively charged.
Each peptide strand consists of 26 amino acids, and the $\alpha$-helix measures approximately 3.8 nm in length. 
In the coiled-coil structures, the helices are approximately 1.2 nm apart.
In total, the system has just under 1000 atoms.
To calculate the rotational grid for coiled-coil peptide and the chlorine anion, 
we placed the center of mass of the coiled-coil peptide at the origin of the coordinate system and aligned the $\alpha$-helices with the $z$-axis.
The relative positions of the chlorine anion Cl$^-$ were generated using the package \texttt{molgri} with a translational grid of 0.50, 0.75, 1.00, 1.25, 1.50, 1.75, 2.00, and 2.25 nm, i.e. an equidistant grid with ~$N_{\mathrm{trans}}=8$.
For the rotational grid of the relative positions, we used an Icosahedron grid with $N_{\mathrm{rot}}=1000$ grid points.
Since Cl$^-$ is a single ion, no rotational grid for the relative orientations was needed.
The combined grid had 8000 grid points and the resulting pseudo-trajectory as many frames.
To give an impression of such a grid, \figref{Closeup}~A shows the relative positions of a Cl$^-$ anion on a similar but more sparse grid with $N_{\mathrm{trans}}=8$ and $N_{\mathrm{rot}}=30$.
We next calculated the interaction energy between Cl$^-$ and the coiled-coil peptide using GROMACS rerun on the pseudo-trajectory.
Since no dynamics are present, we cannot estimate kinetic energy; the interaction energy that we obtain consists of the Coulomb and Lennard-Jones interactions.
Since the coiled-coil peptide is rod-like and the set of Cl$^-$ configurations spherical, some of the structures at smallest radii inevitably feature the chlorine anion overlapping with the peptide, leading to enormous Lennard-Jones interaction energies. We assume that many of these frames can be discarded as unphysical cases.
To obtain more realistic grid point energies, the internal degrees of freedom of the coiled-coil peptide can be relaxed to the nearest local minimum. 
We thus ran a steepest descent energy minimization using GROMACS on every frame of the pseudo-trajectory while constraining the position of the Cl$^-$ anion and the peptide backbone atoms to make sure that the system does not relax into the neighboring grid point.
In three of the 8000 grid points, the energy minimization did not lower the energy, despite very high initial energies. We excluded these grid points from the analysis.
The energy decrease for the other grid points is shown in \figref{Closeup}~B as absolute (black and red line) and relative (blue line) energy difference.
On average, the energy after minimisation is (2200$\pm$ 480) kJ/mol per grid point. 
The changes in energy after minimisation can be roughly divided into three categories, labeled I, II and III in \figref{Closeup}~B and C.
The largest energy differences (I) occur for grid points at which the Cl$^-$ ion overlaps with several atoms of the coiled-coil peptide, usually because it is placed inside the coiled-coil. 
Moderate energy changes (II) occur when the Cl$^-$ overlaps with a single side-chain. 
In the example in the \figref{Closeup}~C-II Cl$^-$ is initially placed on top of a lysine side chain which is located at the exterior of the coiled-coil and can easily rotate away during energy minimization.
In roughly half of the grid points, Cl$^-$  is placed outside of the coiled-coil peptide and only minimal adjustments are made during the energy minimization (III).

We draw the conclusion that minimization is crucial in the areas where Lennard-Jones interaction, especially its repulsive part, is the dominant contribution to energy. Structures from region I and II benefit from subsequent optimization while its effect in zone III is minimal. We can imagine that in docking studies, for example, \texttt{molgri} approach can be used on its own for electrostatically-lead process of the ligand approaching the protein while the subsequent process of the ligand docking within the protein structure can be first studied on a grid and then followed by short optimisation runs.

\begin{figure}[h!]
        \centering
        \includegraphics[width=7cm]{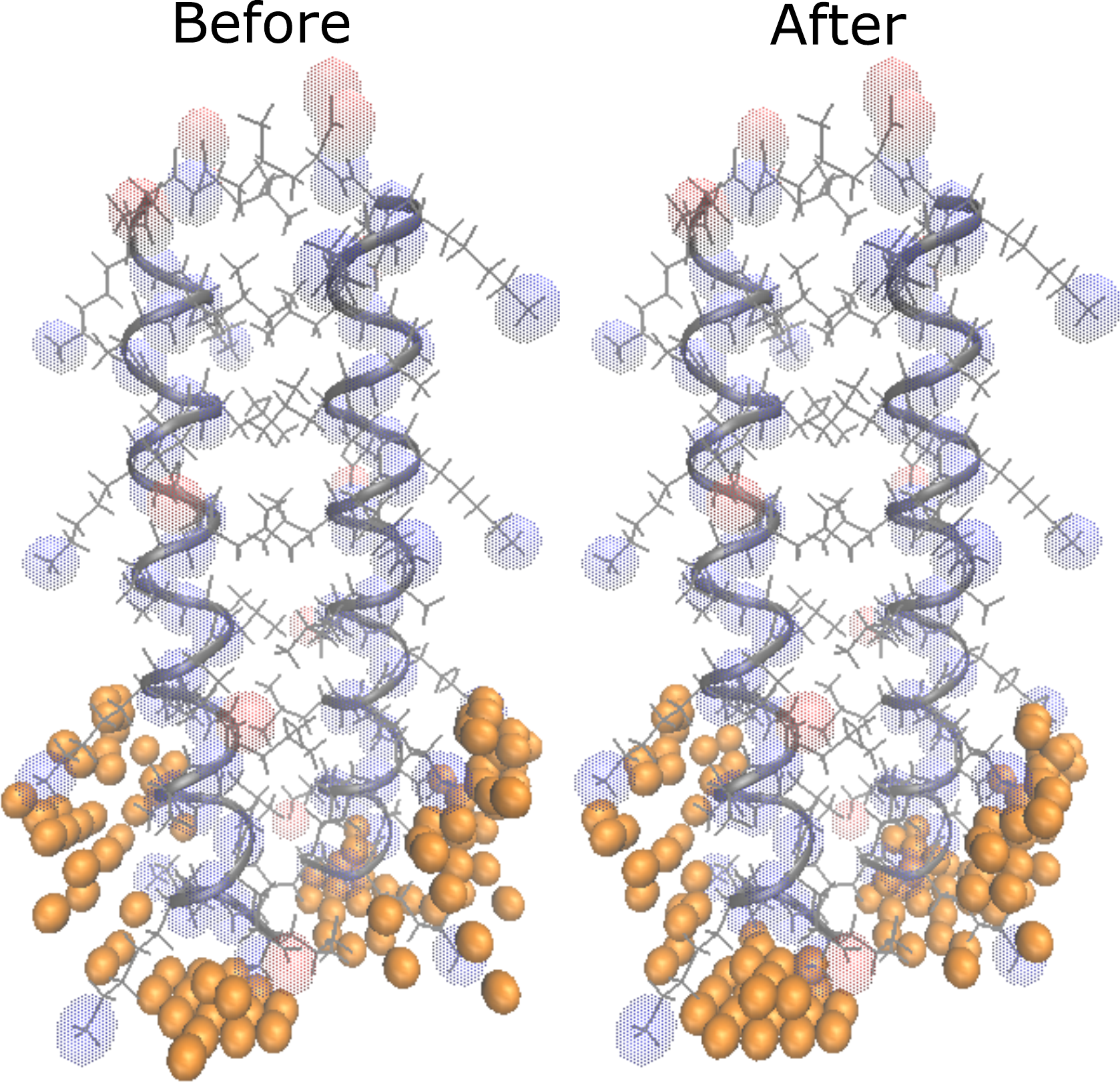}
        \caption{100 frames of coiled-coil-Cl$^-$ pseudo-trajectory ($N_{\mathrm{rot}}=1000$, $N_{\mathrm{trans}}=8$, distances 0.5 - 2.25 nm) with lowest potential energies before (left) and after energy minimization (right). The protein structure (grey) is shown in the initial (non-minimized) form in both figures, Cl$^-$ anions are shown in orange. The red dotted spheres represent negatively charged carboxyl groups and the blue ones represent positively charged amino groups.
        }
        \label{fig:ResultCl}
\end{figure}

%
Since we are using a single anion as our association particle, we can regard it as an electric field probe and expect it to identify positively charged regions of state space.
\figref{ResultCl} shows 100 frames of coiled-coil-Cl$^-$ pseudo-trajectory that had lowest potential energies (before and after minimisation).
We indeed observe that electrostatic interaction is the driving force of association since all low energy structures feature anions congregating around the positively charged N-terminus or the amino groups of the lysine side chains. In addition, changes to the lowest-energy structures with minimization are small. This supports our hypothesis that \texttt{molgri} on its own is a suitable tool to identify regions of interest (those featuring an appropriate charge and a free approach vector) in association studies while energy minimization can be used to fine tune the results. 

\section{Computational Methods}

To test the pseudo-trajectories we generated with \texttt{molgri} we used GROMACS 2022 \cite{van2005gromacs, abraham2015gromacs, lindahl2001gromacs, berendsen1995gromacs} \texttt{rerun} function to perform point energy calculations of configurations proposed by our algorithm. We also used GROMACS to perform comparison MD simulations. Plots of molecular structures were obtained with VMD 1.9 \cite{humphrey1996vmd}. Some figures were combined and labelled in Inkscape 1.0 \cite{harrington2004inkscape}.

For comparison calculations, we used GROMACS simulations with geometry optimisation (without the \texttt{rerun} option). For the water-water example system, we used the rigid TIP3P \cite{jorgensen1983comparison} model of water as a force field and performed simulations with 10 fs time steps. In the polypeptide example systems (hFF03-Cl$^-$) we used the Amber ff99SB-ILDN protein force field \cite{lindorff2010improved}, dielectric constant of water 78.4 (at 300K). All simulations were performed at 300 K with particle mesh Ewald (PME) electrostatics and various trajectory lengths (usually up to 10$^5$ time steps). All parameters were kept constant between the pseudo-trajectory calculation and comparison MD run, although some parameters were irrelevant for a rerun (e.g. time step).

GROMACS was also used for the frame-wise minimisation through the steepest descent algorithm. During this process, the backbone and ion where restrained with GROMACS restraint function, which applies a symmetric force on the chosen atoms. This approach allows changes to the local peptide structure without deforming the whole structure or landing in a neighbouring grid point. To encourage the adaptation of the backbone over ion movement we applied position restraints of 1.000.000 kJ/mol~nm² to Cl$^-$ and 1000 kJ/mol~nm² to protein backbone atoms.
Due to program limitations, the pseudo-trajectory needs to be cut apart, each frame minimised individually and assembled together again.


\section{Conclusion}
We offer a fresh perspective on state space exploration for molecular binding processes by replacing force-field-based sampling of the two-body interaction with a regular grid.
We developed a python workflow that generates relative positions and orientations of two molecules on a regular trans-rot-rot grid, 
and made the program, including ready-made scripts, available as the Python package  \texttt{molgri}. 
%
%
%

%
The quality of a grid-based approach depends heavily on the selection of grid points. To describe a binding process, we used a mesh of translation and rotation grids.
For translations, we employed a linear discretisation of the radial distance between the two molecules.
For rotations, we implemented six different algorithms and compared them using a quantitative uniformity measure.
Algorithms based on polyhedra (Cube 3D and Icosahedron) yielded by far the most uniform rotational grids.
To us this was surprising, because given that the quaternions most naturally describe rotations, we expected quaternion-based algorithms to outperform the other algorithms.
By contrast, algorithms based on Euler angles missed some of the known minima of the water dimer, even with high numbers of grid points, and should only be used with caution for molecular systems.
In terms of computational costs, polyhedra-based algorithms showed the poorest scaling with increasing number of grid points. 
However, since the run time to generate large grids with up to 1000 points is below 100~s on a standard workstation, the computational costs for grid generation is negligible compared to the calculation of grid point energies.
Moreover, rotational grids are generated once and can be re-used for any molecular system. 
Grid points in state space are converted into corresponding molecular structures and stored in a MD-trajectory file format, so that trans-rot-rot grids can easily be interfaced with standard MD programs.
The calculation of the grid point energies can the thus be accomplished by already existing, highly optimised programs.
%
%
The entire method was first applied to two approaching water molecules, showing that the position and depth of potential energy minima found by \texttt{molgri} in combination with polyhedron- or quaternion-based rotational grids matched those found by a stochastic sampling approach.
Moreover, we studied the radial and orientational dependency of energy contributions, reproducing known oriented hydrogen bonding patterns for this system and the dependencies of Lennard-Jones and Coulomb energies on interaction distance. 
While this is not a new result for this simple system, it demonstrates how the untangling of energy contributions on a grid could be performed for more complex systems.
A drawback of our current approach is that it relies on the rigid body approximation, which may be valid for small molecules but is insufficient for induced fit interactions that are commonly encountered in macromolecular systems. 
This problem can be tackled by replacing point energy calculations with short energy minimisations.
We demonstrated this on the system of a coiled-coil peptide binding with a Cl$^-$ anion. %
The GROMACS energy minimzation routine could handle starting structures with strongly overlapping Lennard-Jones spheres, and in most cases the Cl$^-$ position could be accommodated by the rotation of a side chain or a local adjustment of the backbone.
This example also showed that, at large distances, the long-range electrostatic interactions are well-represented by the single-point energy calculations and do not need an energy minimization.
One can further improve the accuracy of the grid point energies by running a short constraint simulation and by including solvent molecules in the energy calculation.
%
%
%
One could further improve the accuracy of the grid point energies by running a short constraint simulation and by including solvent molecules in the energy calculation.
We view the \texttt{molgri} package as a tool to represent the binding process on a grid with approximate energies, to guide the choice of initial structures for unconstrained sampling, or to guide the placement of umbrella potentials for the construction of free-energy surfaces. 
With these use cases in mind, the package can be extended in the following directions:
($i$) providing other non-linear grids, e.g for cylindrical coordinates,
($ii$) improving the interface with  MD programs, such that \texttt{molgri} seamlessly calls a specific MD program, 
($iii$) extending the interface to quantum chemistry packages, such that grid-point energies can be obtained from electronic-structure calculations, 
($iv$) extending the interface to flexible docking algorithms to improve the handling of induced-fit binding.
We are also interested in extending the package to another use case: a rate-model of the binding process constructed on a trans-rot-rot grid using the square-root approximation of the Fokker-Planck equation \cite{bicout1998electron, donati2018estimation, heida2018convergences, donati2021markov, donati2022review}.
For the square-root approximation, the adjacency matrix of the grid, the high-dimensional volumes of the grid cells, and the area of the interface between two neighboring grid cells are required. 
Thus, full control over the grid is critical. 
On a systematic trans-rot-rot grid, all mentioned parameters could be calculated once and applied to various molecular systems.
Whether or not the grid-based approach could ever outperform sampling approaches, assuming that ergodicity is fulfilled, is difficult to say. 
Two aspects of the grid-based approach boost its computational efficiency: 
First, because grid points are independent, calculation of the grid-point energies can be completely parallelised.
Second, because each grid point is evaluated only once, a grid-based approach avoids re-examining the same minima over and over again.
Additionally, by their very nature, grid-based energy-calculations should interface easily with adaptive resolution techniques \cite{cortes2021adaptive}.
On the other hand, because of the high-dimensionality of conformational space, accurate calculations of the grid point energies will have to rely on short stochastic sampling of the orthogonal degrees of freedom in the foreseeable future like we demonstrated in our minimisation routine.
If the exploration of the conformational space in orthogonal degrees of freedom is slow compared to the degrees of freedom resolved by the grid, the computational efficiency is dictated by the sampling.
However, the ergodicity assumption is a constant worry in MD simulations.
In contrast to sampling approaches, grid-based models of molecular interactions do not need to assume ergodicity.
Even with approximate estimates for the grid point (free) energies, grid-based state exploration is a powerful complement to sampling.

\section{Author statements}

\subsection{Competing interests statement}

The authors declare there are no competing interests.

\subsection{Author contribution statement}
H.Z.: Investigation; Methodology; Formal analysis; Software; Visualization; Writing – original draft; Writing – review \& editing.
F.H.: Investigation; Methodology; Formal analysis; Visualization; Writing – review \& editing.
B.G.K.: Conceptualization; Methodology; Funding acquisition; Project administration; Supervision; Writing – review \& editing.

\subsection{Funding statement}
This research has been funded by Deutsche Forschungsgemeinschaft (DFG) through grant 
SFB 1349 Fluorine-Specific Interactions - Project ID 387284271, Project A05;
SFB 1449 Dynamic Hydrogels at Biointerfaces - Project ID 431232613 - Project C02;
RTG 2473 Bioactive Peptides - Project ID 392923329 - Project C01.
We acknowledge access to high-performance computers via the Zentraleinrichtung für Datenverarbeitung (ZEDAT) of Freie Universität Berlin.

\subsection{Data availability statement}

Input molecular structure files used in this contribution and non-commercial software used to manipulate them are included in our software package \texttt{molgri}. We refer to \secref{using_package} for instructions on package acquisition and use. 

\section{Supporting information}

Detailed pseudo-algorithms and code dependencies for the software implementation described in \secref{software} are provided as supporting information. Documentation and source code are hosted on GitHub (\hyperlink{https://github.com/bkellerlab/molecularRotationalGrids}{https://github.com/bkellerlab/molecularRotationalGrids}).


\bibliography{literature}

\providecommand{\latin}[1]{#1}
\makeatletter
\providecommand{\doi}
  {\begingroup\let\do\@makeother\dospecials
  \catcode`\{=1 \catcode`\}=2 \doi@aux}
\providecommand{\doi@aux}[1]{\endgroup\texttt{#1}}
\makeatother
\providecommand*\mcitethebibliography{\thebibliography}
\csname @ifundefined\endcsname{endmcitethebibliography}
  {\let\endmcitethebibliography\endthebibliography}{}
\begin{mcitethebibliography}{63}
\providecommand*\natexlab[1]{#1}
\providecommand*\mciteSetBstSublistMode[1]{}
\providecommand*\mciteSetBstMaxWidthForm[2]{}
\providecommand*\mciteBstWouldAddEndPuncttrue
  {\def\EndOfBibitem{\unskip.}}
\providecommand*\mciteBstWouldAddEndPunctfalse
  {\let\EndOfBibitem\relax}
\providecommand*\mciteSetBstMidEndSepPunct[3]{}
\providecommand*\mciteSetBstSublistLabelBeginEnd[3]{}
\providecommand*\EndOfBibitem{}
\mciteSetBstSublistMode{f}
\mciteSetBstMaxWidthForm{subitem}{(\alph{mcitesubitemcount})}
\mciteSetBstSublistLabelBeginEnd
  {\mcitemaxwidthsubitemform\space}
  {\relax}
  {\relax}

\bibitem[Buckingham(1991)]{buckingham1991hydrogen}
Buckingham,~A. The hydrogen bond, and the structure and properties of H20 and
  (H20) 2. \emph{Journal of molecular structure} \textbf{1991}, \emph{250},
  111--118\relax
\mciteBstWouldAddEndPuncttrue
\mciteSetBstMidEndSepPunct{\mcitedefaultmidpunct}
{\mcitedefaultendpunct}{\mcitedefaultseppunct}\relax
\EndOfBibitem
\bibitem[Rinne \latin{et~al.}(2014)Rinne, Gekle, and Netz]{rinne2014ion}
Rinne,~K.~F.; Gekle,~S.; Netz,~R.~R. Ion-specific solvation water dynamics:
  single water versus collective water effects. \emph{The Journal of Physical
  Chemistry A} \textbf{2014}, \emph{118}, 11667--11677\relax
\mciteBstWouldAddEndPuncttrue
\mciteSetBstMidEndSepPunct{\mcitedefaultmidpunct}
{\mcitedefaultendpunct}{\mcitedefaultseppunct}\relax
\EndOfBibitem
\bibitem[Reinmuth \latin{et~al.}(2019)Reinmuth, Pramanik, Douglas, Day, and
  Bowman-James]{reinmuth2019structural}
Reinmuth,~M.; Pramanik,~S.; Douglas,~J.~T.; Day,~V.~W.; Bowman-James,~K.
  Structural impact of chelation on phytate, a highly phosphorylated
  biomolecule. \emph{European Journal of Inorganic Chemistry} \textbf{2019},
  \emph{2019}, 1870--1874\relax
\mciteBstWouldAddEndPuncttrue
\mciteSetBstMidEndSepPunct{\mcitedefaultmidpunct}
{\mcitedefaultendpunct}{\mcitedefaultseppunct}\relax
\EndOfBibitem
\bibitem[Plattner and No{\'e}(2015)Plattner, and No{\'e}]{plattner2015protein}
Plattner,~N.; No{\'e},~F. Protein conformational plasticity and complex
  ligand-binding kinetics explored by atomistic simulations and Markov models.
  \emph{Nature communications} \textbf{2015}, \emph{6}, 1--10\relax
\mciteBstWouldAddEndPuncttrue
\mciteSetBstMidEndSepPunct{\mcitedefaultmidpunct}
{\mcitedefaultendpunct}{\mcitedefaultseppunct}\relax
\EndOfBibitem
\bibitem[Casasnovas \latin{et~al.}(2017)Casasnovas, Limongelli, Tiwary,
  Carloni, and Parrinello]{casasnovas2017unbinding}
Casasnovas,~R.; Limongelli,~V.; Tiwary,~P.; Carloni,~P.; Parrinello,~M.
  Unbinding kinetics of a p38 MAP kinase type II inhibitor from metadynamics
  simulations. \emph{Journal of the American Chemical Society} \textbf{2017},
  \emph{139}, 4780--4788\relax
\mciteBstWouldAddEndPuncttrue
\mciteSetBstMidEndSepPunct{\mcitedefaultmidpunct}
{\mcitedefaultendpunct}{\mcitedefaultseppunct}\relax
\EndOfBibitem
\bibitem[Kahler \latin{et~al.}(2020)Kahler, Kamenik, Waibl, Kraml, and
  Liedl]{kahler2020protein}
Kahler,~U.; Kamenik,~A.~S.; Waibl,~F.; Kraml,~J.; Liedl,~K.~R. Protein-protein
  binding as a two-step mechanism: Preselection of encounter poses during the
  binding of BPTI and trypsin. \emph{Biophysical journal} \textbf{2020},
  \emph{119}, 652--666\relax
\mciteBstWouldAddEndPuncttrue
\mciteSetBstMidEndSepPunct{\mcitedefaultmidpunct}
{\mcitedefaultendpunct}{\mcitedefaultseppunct}\relax
\EndOfBibitem
\bibitem[Wenz \latin{et~al.}(2021)Wenz, Bertazzon, Sticht, Aleksic,
  Gjorgjevikj, Freund, and Keller]{wenz2021target}
Wenz,~M.~T.; Bertazzon,~M.; Sticht,~J.; Aleksic,~S.; Gjorgjevikj,~D.;
  Freund,~C.; Keller,~B.~G. Target recognition in tandem WW domains: complex
  structures for parallel and antiparallel ligand orientation in h-FBP21 tandem
  WW. \emph{Journal of Chemical Information and Modeling} \textbf{2021}, \relax
\mciteBstWouldAddEndPunctfalse
\mciteSetBstMidEndSepPunct{\mcitedefaultmidpunct}
{}{\mcitedefaultseppunct}\relax
\EndOfBibitem
\bibitem[Morriss-Andrews and Shea(2015)Morriss-Andrews, and
  Shea]{morriss2015computational}
Morriss-Andrews,~A.; Shea,~J.-E. Computational studies of protein aggregation:
  methods and applications. \emph{Annual review of physical chemistry}
  \textbf{2015}, \emph{66}, 643--666\relax
\mciteBstWouldAddEndPuncttrue
\mciteSetBstMidEndSepPunct{\mcitedefaultmidpunct}
{\mcitedefaultendpunct}{\mcitedefaultseppunct}\relax
\EndOfBibitem
\bibitem[Morozova and Muthukumar(2018)Morozova, and
  Muthukumar]{morozova2018electrostatic}
Morozova,~S.; Muthukumar,~M. Electrostatic effects in collagen fibril
  formation. \emph{The Journal of chemical physics} \textbf{2018}, \emph{149},
  163333\relax
\mciteBstWouldAddEndPuncttrue
\mciteSetBstMidEndSepPunct{\mcitedefaultmidpunct}
{\mcitedefaultendpunct}{\mcitedefaultseppunct}\relax
\EndOfBibitem
\bibitem[Hellmund \latin{et~al.}(2021)Hellmund, Von~Lospichl, B{\"o}ttcher,
  Ludwig, Keiderling, Noirez, Wei{\ss}, Mikolajczak, Gradzielski, and
  Koksch]{hellmund2021functionalized}
Hellmund,~K.~S.; Von~Lospichl,~B.; B{\"o}ttcher,~C.; Ludwig,~K.;
  Keiderling,~U.; Noirez,~L.; Wei{\ss},~A.; Mikolajczak,~D.~J.;
  Gradzielski,~M.; Koksch,~B. Functionalized peptide hydrogels as tunable
  extracellular matrix mimics for biological applications. \emph{Peptide
  Science} \textbf{2021}, \emph{113}, e24201\relax
\mciteBstWouldAddEndPuncttrue
\mciteSetBstMidEndSepPunct{\mcitedefaultmidpunct}
{\mcitedefaultendpunct}{\mcitedefaultseppunct}\relax
\EndOfBibitem
\bibitem[Bolhuis \latin{et~al.}(2002)Bolhuis, Chandler, Dellago, and
  Geissler]{bolhuis2002transition}
Bolhuis,~P.~G.; Chandler,~D.; Dellago,~C.; Geissler,~P.~L. Transition Path
  Sampling: Throwing Ropes. \emph{Annu. Rev. Phys. Chem} \textbf{2002},
  \emph{53}, 291--318\relax
\mciteBstWouldAddEndPuncttrue
\mciteSetBstMidEndSepPunct{\mcitedefaultmidpunct}
{\mcitedefaultendpunct}{\mcitedefaultseppunct}\relax
\EndOfBibitem
\bibitem[Barducci \latin{et~al.}(2011)Barducci, Bonomi, and
  Parrinello]{barducci2011metadynamics}
Barducci,~A.; Bonomi,~M.; Parrinello,~M. Metadynamics. \emph{Wiley
  Interdisciplinary Reviews: Computational Molecular Science} \textbf{2011},
  \emph{1}, 826--843\relax
\mciteBstWouldAddEndPuncttrue
\mciteSetBstMidEndSepPunct{\mcitedefaultmidpunct}
{\mcitedefaultendpunct}{\mcitedefaultseppunct}\relax
\EndOfBibitem
\bibitem[Zuckerman and Chong(2017)Zuckerman, and Chong]{zuckerman2017weighted}
Zuckerman,~D.~M.; Chong,~L.~T. Weighted ensemble simulation: review of
  methodology, applications, and software. \emph{Annual review of biophysics}
  \textbf{2017}, \emph{46}, 43\relax
\mciteBstWouldAddEndPuncttrue
\mciteSetBstMidEndSepPunct{\mcitedefaultmidpunct}
{\mcitedefaultendpunct}{\mcitedefaultseppunct}\relax
\EndOfBibitem
\bibitem[Bowman \latin{et~al.}(2010)Bowman, Ensign, and
  Pande]{bowman2010enhanced}
Bowman,~G.~R.; Ensign,~D.~L.; Pande,~V.~S. Enhanced modeling via network
  theory: Adaptive sampling of Markov state models. \emph{Journal of chemical
  theory and computation} \textbf{2010}, \emph{6}, 787--794\relax
\mciteBstWouldAddEndPuncttrue
\mciteSetBstMidEndSepPunct{\mcitedefaultmidpunct}
{\mcitedefaultendpunct}{\mcitedefaultseppunct}\relax
\EndOfBibitem
\bibitem[Bruce \latin{et~al.}(2018)Bruce, Ganotra, Kokh, Sadiq, and
  Wade]{bruce2018new}
Bruce,~N.~J.; Ganotra,~G.~K.; Kokh,~D.~B.; Sadiq,~S.~K.; Wade,~R.~C. New
  approaches for computing ligand--receptor binding kinetics. \emph{Current
  opinion in structural biology} \textbf{2018}, \emph{49}, 1--10\relax
\mciteBstWouldAddEndPuncttrue
\mciteSetBstMidEndSepPunct{\mcitedefaultmidpunct}
{\mcitedefaultendpunct}{\mcitedefaultseppunct}\relax
\EndOfBibitem
\bibitem[Limongelli(2020)]{limongelli2020ligand}
Limongelli,~V. Ligand binding free energy and kinetics calculation in 2020.
  \emph{Wiley Interdisciplinary Reviews: Computational Molecular Science}
  \textbf{2020}, \emph{10}, e1455\relax
\mciteBstWouldAddEndPuncttrue
\mciteSetBstMidEndSepPunct{\mcitedefaultmidpunct}
{\mcitedefaultendpunct}{\mcitedefaultseppunct}\relax
\EndOfBibitem
\bibitem[Plattner \latin{et~al.}(2017)Plattner, Doerr, De~Fabritiis, and
  No{\'e}]{plattner2017complete}
Plattner,~N.; Doerr,~S.; De~Fabritiis,~G.; No{\'e},~F. Complete
  protein--protein association kinetics in atomic detail revealed by molecular
  dynamics simulations and Markov modelling. \emph{Nature chemistry}
  \textbf{2017}, \emph{9}, 1005--1011\relax
\mciteBstWouldAddEndPuncttrue
\mciteSetBstMidEndSepPunct{\mcitedefaultmidpunct}
{\mcitedefaultendpunct}{\mcitedefaultseppunct}\relax
\EndOfBibitem
\bibitem[Capelli \latin{et~al.}(2019)Capelli, Carloni, and
  Parrinello]{capelli2019exhaustive}
Capelli,~R.; Carloni,~P.; Parrinello,~M. Exhaustive search of ligand binding
  pathways via volume-based metadynamics. \emph{The journal of physical
  chemistry letters} \textbf{2019}, \emph{10}, 3495--3499\relax
\mciteBstWouldAddEndPuncttrue
\mciteSetBstMidEndSepPunct{\mcitedefaultmidpunct}
{\mcitedefaultendpunct}{\mcitedefaultseppunct}\relax
\EndOfBibitem
\bibitem[Lemke \latin{et~al.}(2018)Lemke, Peter, and
  Kukharenko]{lemke2018efficient}
Lemke,~T.; Peter,~C.; Kukharenko,~O. Efficient Sampling and Characterization of
  Free Energy Landscapes of Ion--Peptide Systems. \emph{Journal of chemical
  theory and computation} \textbf{2018}, \emph{14}, 5476--5488\relax
\mciteBstWouldAddEndPuncttrue
\mciteSetBstMidEndSepPunct{\mcitedefaultmidpunct}
{\mcitedefaultendpunct}{\mcitedefaultseppunct}\relax
\EndOfBibitem
\bibitem[Batra \latin{et~al.}(2013)Batra, Szab{\'o}, Caulfield, Soares,
  Sahin-T{\'o}th, and Radisky]{batra2013long}
Batra,~J.; Szab{\'o},~A.; Caulfield,~T.~R.; Soares,~A.~S.; Sahin-T{\'o}th,~M.;
  Radisky,~E.~S. Long-range electrostatic complementarity governs substrate
  recognition by human chymotrypsin C, a key regulator of digestive enzyme
  activation. \emph{Journal of Biological Chemistry} \textbf{2013}, \emph{288},
  9848--9859\relax
\mciteBstWouldAddEndPuncttrue
\mciteSetBstMidEndSepPunct{\mcitedefaultmidpunct}
{\mcitedefaultendpunct}{\mcitedefaultseppunct}\relax
\EndOfBibitem
\bibitem[Waldner \latin{et~al.}(2018)Waldner, Kraml, Kahler, Spinn, Schauperl,
  Podewitz, Fuchs, Cruciani, and Liedl]{waldner2018electrostatic}
Waldner,~B.~J.; Kraml,~J.; Kahler,~U.; Spinn,~A.; Schauperl,~M.; Podewitz,~M.;
  Fuchs,~J.~E.; Cruciani,~G.; Liedl,~K.~R. Electrostatic recognition in
  substrate binding to serine proteases. \emph{Journal of Molecular
  Recognition} \textbf{2018}, \emph{31}, e2727\relax
\mciteBstWouldAddEndPuncttrue
\mciteSetBstMidEndSepPunct{\mcitedefaultmidpunct}
{\mcitedefaultendpunct}{\mcitedefaultseppunct}\relax
\EndOfBibitem
\bibitem[Chodera(2016)]{chodera2016simple}
Chodera,~J.~D. A simple method for automated equilibration detection in
  molecular simulations. \emph{Journal of chemical theory and computation}
  \textbf{2016}, \emph{12}, 1799--1805\relax
\mciteBstWouldAddEndPuncttrue
\mciteSetBstMidEndSepPunct{\mcitedefaultmidpunct}
{\mcitedefaultendpunct}{\mcitedefaultseppunct}\relax
\EndOfBibitem
\bibitem[Bicout and Szabo(1998)Bicout, and Szabo]{bicout1998electron}
Bicout,~D.; Szabo,~A. Electron transfer reaction dynamics in non-Debye
  solvents. \emph{The Journal of chemical physics} \textbf{1998}, \emph{109},
  2325--2338\relax
\mciteBstWouldAddEndPuncttrue
\mciteSetBstMidEndSepPunct{\mcitedefaultmidpunct}
{\mcitedefaultendpunct}{\mcitedefaultseppunct}\relax
\EndOfBibitem
\bibitem[Donati \latin{et~al.}(2018)Donati, Heida, Keller, and
  Weber]{donati2018estimation}
Donati,~L.; Heida,~M.; Keller,~B.~G.; Weber,~M. Estimation of the infinitesimal
  generator by square-root approximation. \emph{Journal of Physics: Condensed
  Matter} \textbf{2018}, \emph{30}, 425201\relax
\mciteBstWouldAddEndPuncttrue
\mciteSetBstMidEndSepPunct{\mcitedefaultmidpunct}
{\mcitedefaultendpunct}{\mcitedefaultseppunct}\relax
\EndOfBibitem
\bibitem[Heida(2018)]{heida2018convergences}
Heida,~M. Convergences of the squareroot approximation scheme to the
  Fokker--Planck operator. \emph{Mathematical Models and Methods in Applied
  Sciences} \textbf{2018}, \emph{28}, 2599--2635\relax
\mciteBstWouldAddEndPuncttrue
\mciteSetBstMidEndSepPunct{\mcitedefaultmidpunct}
{\mcitedefaultendpunct}{\mcitedefaultseppunct}\relax
\EndOfBibitem
\bibitem[Donati \latin{et~al.}(2021)Donati, Weber, and
  Keller]{donati2021markov}
Donati,~L.; Weber,~M.; Keller,~B.~G. Markov models from the square root
  approximation of the Fokker--Planck equation: calculating the grid-dependent
  flux. \emph{Journal of Physics: Condensed Matter} \textbf{2021}, \emph{33},
  115902\relax
\mciteBstWouldAddEndPuncttrue
\mciteSetBstMidEndSepPunct{\mcitedefaultmidpunct}
{\mcitedefaultendpunct}{\mcitedefaultseppunct}\relax
\EndOfBibitem
\bibitem[Donati \latin{et~al.}(2022)Donati, Weber, and
  Keller]{donati2022review}
Donati,~L.; Weber,~M.; Keller,~B.~G. A review of Girsanov Reweighting and of
  Square Root Approximation for building molecular Markov State Models.
  \emph{arXiv preprint arXiv:2209.10544} \textbf{2022}, \relax
\mciteBstWouldAddEndPunctfalse
\mciteSetBstMidEndSepPunct{\mcitedefaultmidpunct}
{}{\mcitedefaultseppunct}\relax
\EndOfBibitem
\bibitem[Euler(1776)]{euler1776nova}
Euler,~L. Nova methodus motum corporum rigidorum degerminandi. \emph{Novi
  commentarii academiae scientiarum Petropolitanae} \textbf{1776},
  208--238\relax
\mciteBstWouldAddEndPuncttrue
\mciteSetBstMidEndSepPunct{\mcitedefaultmidpunct}
{\mcitedefaultendpunct}{\mcitedefaultseppunct}\relax
\EndOfBibitem
\bibitem[Kirk(2012)]{kirk2012graphics}
Kirk,~D. \emph{Graphics Gems III (IBM Version)}; Elsevier, 2012\relax
\mciteBstWouldAddEndPuncttrue
\mciteSetBstMidEndSepPunct{\mcitedefaultmidpunct}
{\mcitedefaultendpunct}{\mcitedefaultseppunct}\relax
\EndOfBibitem
\bibitem[Yershova and LaValle(2004)Yershova, and
  LaValle]{yershova2004deterministic}
Yershova,~A.; LaValle,~S.~M. Deterministic sampling methods for spheres and SO
  (3). IEEE International Conference on Robotics and Automation. 2004; pp
  3974--3980\relax
\mciteBstWouldAddEndPuncttrue
\mciteSetBstMidEndSepPunct{\mcitedefaultmidpunct}
{\mcitedefaultendpunct}{\mcitedefaultseppunct}\relax
\EndOfBibitem
\bibitem[Yershova \latin{et~al.}(2010)Yershova, Jain, Lavalle, and
  Mitchell]{yershova2010generating}
Yershova,~A.; Jain,~S.; Lavalle,~S.~M.; Mitchell,~J.~C. Generating uniform
  incremental grids on SO (3) using the Hopf fibration. \emph{The International
  journal of robotics research} \textbf{2010}, \emph{29}, 801--812\relax
\mciteBstWouldAddEndPuncttrue
\mciteSetBstMidEndSepPunct{\mcitedefaultmidpunct}
{\mcitedefaultendpunct}{\mcitedefaultseppunct}\relax
\EndOfBibitem
\bibitem[Karney(2007)]{karney2007quaternions}
Karney,~C.~F. Quaternions in molecular modeling. \emph{Journal of Molecular
  Graphics and Modelling} \textbf{2007}, \emph{25}, 595--604\relax
\mciteBstWouldAddEndPuncttrue
\mciteSetBstMidEndSepPunct{\mcitedefaultmidpunct}
{\mcitedefaultendpunct}{\mcitedefaultseppunct}\relax
\EndOfBibitem
\bibitem[Diebel(2006)]{diebel2006representing}
Diebel,~J. Representing attitude: Euler angles, unit quaternions, and rotation
  vectors. \emph{Matrix} \textbf{2006}, \emph{58}, 1--35\relax
\mciteBstWouldAddEndPuncttrue
\mciteSetBstMidEndSepPunct{\mcitedefaultmidpunct}
{\mcitedefaultendpunct}{\mcitedefaultseppunct}\relax
\EndOfBibitem
\bibitem[Purser and Ran{\v{c}}i{\'c}(2011)Purser, and
  Ran{\v{c}}i{\'c}]{purser2011standardized}
Purser,~R.~J.; Ran{\v{c}}i{\'c},~M. \emph{A standardized procedure for the
  derivation of smooth and partially overset grids on the sphere, associated
  with polyhedra that admit regular griddings of their surfaces; Mathematical
  principles of classification and construction; Part I}; 2011\relax
\mciteBstWouldAddEndPuncttrue
\mciteSetBstMidEndSepPunct{\mcitedefaultmidpunct}
{\mcitedefaultendpunct}{\mcitedefaultseppunct}\relax
\EndOfBibitem
\bibitem[Sadourny(1972)]{sadourny1972conservative}
Sadourny,~R. Conservative finite-difference approximations of the primitive
  equations on quasi-uniform spherical grids. \emph{Monthly Weather Review}
  \textbf{1972}, \emph{100}, 136--144\relax
\mciteBstWouldAddEndPuncttrue
\mciteSetBstMidEndSepPunct{\mcitedefaultmidpunct}
{\mcitedefaultendpunct}{\mcitedefaultseppunct}\relax
\EndOfBibitem
\bibitem[Lindemann \latin{et~al.}(2004)Lindemann, Yershova, and
  LaValle]{lindemann2004incremental}
Lindemann,~S.~R.; Yershova,~A.; LaValle,~S.~M. \emph{Algorithmic Foundations of
  Robotics VI}; Springer, 2004; pp 313--328\relax
\mciteBstWouldAddEndPuncttrue
\mciteSetBstMidEndSepPunct{\mcitedefaultmidpunct}
{\mcitedefaultendpunct}{\mcitedefaultseppunct}\relax
\EndOfBibitem
\bibitem[Kneller(1991)]{kneller1991superposition}
Kneller,~G.~R. Superposition of molecular structures using quaternions.
  \emph{Molecular Simulation} \textbf{1991}, \emph{7}, 113--119\relax
\mciteBstWouldAddEndPuncttrue
\mciteSetBstMidEndSepPunct{\mcitedefaultmidpunct}
{\mcitedefaultendpunct}{\mcitedefaultseppunct}\relax
\EndOfBibitem
\bibitem[Fincham(1992)]{fincham1992leapfrog}
Fincham,~D. Leapfrog rotational algorithms. \emph{Molecular Simulation}
  \textbf{1992}, \emph{8}, 165--178\relax
\mciteBstWouldAddEndPuncttrue
\mciteSetBstMidEndSepPunct{\mcitedefaultmidpunct}
{\mcitedefaultendpunct}{\mcitedefaultseppunct}\relax
\EndOfBibitem
\bibitem[Kol \latin{et~al.}(1997)Kol, Laird, and Leimkuhler]{kol1997symplectic}
Kol,~A.; Laird,~B.~B.; Leimkuhler,~B.~J. A symplectic method for rigid-body
  molecular simulation. \emph{The Journal of chemical physics} \textbf{1997},
  \emph{107}, 2580--2588\relax
\mciteBstWouldAddEndPuncttrue
\mciteSetBstMidEndSepPunct{\mcitedefaultmidpunct}
{\mcitedefaultendpunct}{\mcitedefaultseppunct}\relax
\EndOfBibitem
\bibitem[Miller~Iii \latin{et~al.}(2002)Miller~Iii, Eleftheriou, Pattnaik,
  Ndirango, Newns, and Martyna]{miller2002symplectic}
Miller~Iii,~T.; Eleftheriou,~M.; Pattnaik,~P.; Ndirango,~A.; Newns,~D.;
  Martyna,~G. Symplectic quaternion scheme for biophysical molecular dynamics.
  \emph{The Journal of chemical physics} \textbf{2002}, \emph{116},
  8649--8659\relax
\mciteBstWouldAddEndPuncttrue
\mciteSetBstMidEndSepPunct{\mcitedefaultmidpunct}
{\mcitedefaultendpunct}{\mcitedefaultseppunct}\relax
\EndOfBibitem
\bibitem[Nielsen \latin{et~al.}(2010)Nielsen, Bulo, Moore, and
  Ensing]{nielsen2010recent}
Nielsen,~S.~O.; Bulo,~R.~E.; Moore,~P.~B.; Ensing,~B. Recent progress in
  adaptive multiscale molecular dynamics simulations of soft matter.
  \emph{Physical Chemistry Chemical Physics} \textbf{2010}, \emph{12},
  12401--12414\relax
\mciteBstWouldAddEndPuncttrue
\mciteSetBstMidEndSepPunct{\mcitedefaultmidpunct}
{\mcitedefaultendpunct}{\mcitedefaultseppunct}\relax
\EndOfBibitem
\bibitem[Stumpe and Grubm{\"u}ller(2007)Stumpe, and
  Grubm{\"u}ller]{stumpe2007aqueous}
Stumpe,~M.~C.; Grubm{\"u}ller,~H. Aqueous urea solutions: structure,
  energetics, and urea aggregation. \emph{The Journal of Physical Chemistry B}
  \textbf{2007}, \emph{111}, 6220--6228\relax
\mciteBstWouldAddEndPuncttrue
\mciteSetBstMidEndSepPunct{\mcitedefaultmidpunct}
{\mcitedefaultendpunct}{\mcitedefaultseppunct}\relax
\EndOfBibitem
\bibitem[Heinz and Grubm{\"u}ller(2019)Heinz, and
  Grubm{\"u}ller]{heinz2019computing}
Heinz,~L.~P.; Grubm{\"u}ller,~H. Computing spatially resolved rotational
  hydration entropies from atomistic simulations. \emph{Journal of Chemical
  Theory and Computation} \textbf{2019}, \emph{16}, 108--118\relax
\mciteBstWouldAddEndPuncttrue
\mciteSetBstMidEndSepPunct{\mcitedefaultmidpunct}
{\mcitedefaultendpunct}{\mcitedefaultseppunct}\relax
\EndOfBibitem
\bibitem[Lynden-Bell and Stone(1989)Lynden-Bell, and
  Stone]{lynden1989reorientational}
Lynden-Bell,~R.; Stone,~A. Reorientational correlation functions, quaternions
  and Wigner rotation matrices. \emph{Molecular Simulation} \textbf{1989},
  \emph{3}, 271--281\relax
\mciteBstWouldAddEndPuncttrue
\mciteSetBstMidEndSepPunct{\mcitedefaultmidpunct}
{\mcitedefaultendpunct}{\mcitedefaultseppunct}\relax
\EndOfBibitem
\bibitem[{Wikipedia contributors}(2022)]{wiki_rot_matrix}
{Wikipedia contributors}, Rotation matrix - {W}ikipedia{,} The Free
  Encyclopedia. 2022;
  \url{https://en.wikipedia.org/wiki/Rotation_matrix#In_three_dimensions},
  [accessed 24-October-2022]\relax
\mciteBstWouldAddEndPuncttrue
\mciteSetBstMidEndSepPunct{\mcitedefaultmidpunct}
{\mcitedefaultendpunct}{\mcitedefaultseppunct}\relax
\EndOfBibitem
\bibitem[Hamilton(1840)]{hamilton1840new}
Hamilton,~W.~R. On a new species of imaginary quantities, connected with the
  theory of quaternions. \emph{Proceedings of the Royal Irish Academy
  (1836-1869)} \textbf{1840}, \emph{2}, 424--434\relax
\mciteBstWouldAddEndPuncttrue
\mciteSetBstMidEndSepPunct{\mcitedefaultmidpunct}
{\mcitedefaultendpunct}{\mcitedefaultseppunct}\relax
\EndOfBibitem
\bibitem[Hamilton(1850)]{hamilton1850quaternions}
Hamilton,~W.~R. On quaternions and the rotation of a solid body. Proceedings of
  the Royal Irish Academy. 1850; pp 38--56\relax
\mciteBstWouldAddEndPuncttrue
\mciteSetBstMidEndSepPunct{\mcitedefaultmidpunct}
{\mcitedefaultendpunct}{\mcitedefaultseppunct}\relax
\EndOfBibitem
\bibitem[Evlero and Petropolitanae(1770)Evlero, and
  Petropolitanae]{evlero1770problema}
Evlero,~A.~L.; Petropolitanae,~N. Problema algebraicvm ob affectiones prorsvs
  singvlares memorabile. \emph{Novi Commentarii Academiae imperialis
  scientiarum Petropolitanae} \textbf{1770}, \relax
\mciteBstWouldAddEndPunctfalse
\mciteSetBstMidEndSepPunct{\mcitedefaultmidpunct}
{}{\mcitedefaultseppunct}\relax
\EndOfBibitem
\bibitem[Rodrigues(1840)]{rodrigues1840lois}
Rodrigues,~O. Des lois g{\'e}om{\'e}triques qui r{\'e}gissent les
  d{\'e}placements d’un syst{\`e}me solide dans l’espace, et de la
  variation des coordonn{\'e}es provenant de ces d{\'e}placements
  consid{\'e}r{\'e}s ind{\'e}pendamment des causes qui peuvent les produire.
  \emph{J. Math. Pures Appl} \textbf{1840}, \emph{5}, 5\relax
\mciteBstWouldAddEndPuncttrue
\mciteSetBstMidEndSepPunct{\mcitedefaultmidpunct}
{\mcitedefaultendpunct}{\mcitedefaultseppunct}\relax
\EndOfBibitem
\bibitem[Lynch and Park(2017)Lynch, and Park]{lynch2017modern}
Lynch,~K.~M.; Park,~F.~C. \emph{Modern robotics}; Cambridge University Press,
  2017\relax
\mciteBstWouldAddEndPuncttrue
\mciteSetBstMidEndSepPunct{\mcitedefaultmidpunct}
{\mcitedefaultendpunct}{\mcitedefaultseppunct}\relax
\EndOfBibitem
\bibitem[{Wikipedia contributors}(2022)]{wiki_spherical_cap}
{Wikipedia contributors}, Spherical cap --- {W}ikipedia{,} The Free
  Encyclopedia. 2022; \url{{https://en.wikipedia.org/wiki/Spherical_cap}},
  [accessed 24-October-2022]\relax
\mciteBstWouldAddEndPuncttrue
\mciteSetBstMidEndSepPunct{\mcitedefaultmidpunct}
{\mcitedefaultendpunct}{\mcitedefaultseppunct}\relax
\EndOfBibitem
\bibitem[mol(2022)]{molgri_pypi}
pypi package \texttt{molgri}. 2022; \url{https://pypi.org/project/molgri/},
  [accessed 29-October-2022]\relax
\mciteBstWouldAddEndPuncttrue
\mciteSetBstMidEndSepPunct{\mcitedefaultmidpunct}
{\mcitedefaultendpunct}{\mcitedefaultseppunct}\relax
\EndOfBibitem
\bibitem[mol(2022)]{molgri_GitHub}
GitHub repository \texttt{molgri}. 2022;
  \url{https://github.com/bkellerlab/molecularRotationalGrids}, [accessed
  29-October-2022]\relax
\mciteBstWouldAddEndPuncttrue
\mciteSetBstMidEndSepPunct{\mcitedefaultmidpunct}
{\mcitedefaultendpunct}{\mcitedefaultseppunct}\relax
\EndOfBibitem
\bibitem[Woolfson(2005)]{woolfson2005design}
Woolfson,~D.~N. The design of coiled-coil structures and assemblies.
  \emph{Advances in protein chemistry} \textbf{2005}, \emph{70}, 79--112\relax
\mciteBstWouldAddEndPuncttrue
\mciteSetBstMidEndSepPunct{\mcitedefaultmidpunct}
{\mcitedefaultendpunct}{\mcitedefaultseppunct}\relax
\EndOfBibitem
\bibitem[Van Der~Spoel \latin{et~al.}(2005)Van Der~Spoel, Lindahl, Hess,
  Groenhof, Mark, and Berendsen]{van2005gromacs}
Van Der~Spoel,~D.; Lindahl,~E.; Hess,~B.; Groenhof,~G.; Mark,~A.~E.;
  Berendsen,~H.~J. GROMACS: fast, flexible, and free. \emph{Journal of
  computational chemistry} \textbf{2005}, \emph{26}, 1701--1718\relax
\mciteBstWouldAddEndPuncttrue
\mciteSetBstMidEndSepPunct{\mcitedefaultmidpunct}
{\mcitedefaultendpunct}{\mcitedefaultseppunct}\relax
\EndOfBibitem
\bibitem[Abraham \latin{et~al.}(2015)Abraham, Murtola, Schulz, P{\'a}ll, Smith,
  Hess, and Lindahl]{abraham2015gromacs}
Abraham,~M.~J.; Murtola,~T.; Schulz,~R.; P{\'a}ll,~S.; Smith,~J.~C.; Hess,~B.;
  Lindahl,~E. GROMACS: High performance molecular simulations through
  multi-level parallelism from laptops to supercomputers. \emph{SoftwareX}
  \textbf{2015}, \emph{1}, 19--25\relax
\mciteBstWouldAddEndPuncttrue
\mciteSetBstMidEndSepPunct{\mcitedefaultmidpunct}
{\mcitedefaultendpunct}{\mcitedefaultseppunct}\relax
\EndOfBibitem
\bibitem[Lindahl \latin{et~al.}(2001)Lindahl, Hess, and Van
  Der~Spoel]{lindahl2001gromacs}
Lindahl,~E.; Hess,~B.; Van Der~Spoel,~D. GROMACS 3.0: a package for molecular
  simulation and trajectory analysis. \emph{Molecular modeling annual}
  \textbf{2001}, \emph{7}, 306--317\relax
\mciteBstWouldAddEndPuncttrue
\mciteSetBstMidEndSepPunct{\mcitedefaultmidpunct}
{\mcitedefaultendpunct}{\mcitedefaultseppunct}\relax
\EndOfBibitem
\bibitem[Berendsen \latin{et~al.}(1995)Berendsen, van~der Spoel, and van
  Drunen]{berendsen1995gromacs}
Berendsen,~H.~J.; van~der Spoel,~D.; van Drunen,~R. GROMACS: A message-passing
  parallel molecular dynamics implementation. \emph{Computer physics
  communications} \textbf{1995}, \emph{91}, 43--56\relax
\mciteBstWouldAddEndPuncttrue
\mciteSetBstMidEndSepPunct{\mcitedefaultmidpunct}
{\mcitedefaultendpunct}{\mcitedefaultseppunct}\relax
\EndOfBibitem
\bibitem[Humphrey \latin{et~al.}(1996)Humphrey, Dalke, and
  Schulten]{humphrey1996vmd}
Humphrey,~W.; Dalke,~A.; Schulten,~K. VMD: visual molecular dynamics.
  \emph{Journal of molecular graphics} \textbf{1996}, \emph{14}, 33--38\relax
\mciteBstWouldAddEndPuncttrue
\mciteSetBstMidEndSepPunct{\mcitedefaultmidpunct}
{\mcitedefaultendpunct}{\mcitedefaultseppunct}\relax
\EndOfBibitem
\bibitem[Harrington and Engelen(2004)Harrington, and
  Engelen]{harrington2004inkscape}
Harrington,~B.; Engelen,~J. Inkscape. 2004; Software available at http://www.
  inkscape. org\relax
\mciteBstWouldAddEndPuncttrue
\mciteSetBstMidEndSepPunct{\mcitedefaultmidpunct}
{\mcitedefaultendpunct}{\mcitedefaultseppunct}\relax
\EndOfBibitem
\bibitem[Jorgensen \latin{et~al.}(1983)Jorgensen, Chandrasekhar, Madura, Impey,
  and Klein]{jorgensen1983comparison}
Jorgensen,~W.~L.; Chandrasekhar,~J.; Madura,~J.~D.; Impey,~R.~W.; Klein,~M.~L.
  Comparison of simple potential functions for simulating liquid water.
  \emph{The Journal of chemical physics} \textbf{1983}, \emph{79},
  926--935\relax
\mciteBstWouldAddEndPuncttrue
\mciteSetBstMidEndSepPunct{\mcitedefaultmidpunct}
{\mcitedefaultendpunct}{\mcitedefaultseppunct}\relax
\EndOfBibitem
\bibitem[Lindorff-Larsen \latin{et~al.}(2010)Lindorff-Larsen, Piana, Palmo,
  Maragakis, Klepeis, Dror, and Shaw]{lindorff2010improved}
Lindorff-Larsen,~K.; Piana,~S.; Palmo,~K.; Maragakis,~P.; Klepeis,~J.~L.;
  Dror,~R.~O.; Shaw,~D.~E. Improved side-chain torsion potentials for the Amber
  ff99SB protein force field. \emph{Proteins: Structure, Function, and
  Bioinformatics} \textbf{2010}, \emph{78}, 1950--1958\relax
\mciteBstWouldAddEndPuncttrue
\mciteSetBstMidEndSepPunct{\mcitedefaultmidpunct}
{\mcitedefaultendpunct}{\mcitedefaultseppunct}\relax
\EndOfBibitem
\bibitem[Cortes-Huerto \latin{et~al.}(2021)Cortes-Huerto, Praprotnik, Kremer,
  and Delle~Site]{cortes2021adaptive}
Cortes-Huerto,~R.; Praprotnik,~M.; Kremer,~K.; Delle~Site,~L. From adaptive
  resolution to molecular dynamics of open systems. \emph{The European Physical
  Journal B} \textbf{2021}, \emph{94}, 1--22\relax
\mciteBstWouldAddEndPuncttrue
\mciteSetBstMidEndSepPunct{\mcitedefaultmidpunct}
{\mcitedefaultendpunct}{\mcitedefaultseppunct}\relax
\EndOfBibitem
\end{mcitethebibliography}

\end{document}


\preprint{AIP/123-QED}

\title[SI: Grid-based state space exploration]{Supplementary Information: Grid-based state space exploration for molecular binding}
\author{Hana Zupan}
\author{Frederick Heinz}
\author{Bettina G.~Keller}%

\date{\today}

%
%
\begin{abstract}

\end{abstract}

\maketitle


\section{Dependencies of the \texttt{molgri}-package}
\label{app:dependencies}
The \texttt{molgri} package is written entirely in Python 3.9 \cite{ref_python} using standard scientific programming libraries as dependencies: numpy 1.23.0 \cite{harris2020array} for fast calculations with arrays, scipy 1.8.1 \cite{2020SciPy-NMeth} for distance algorithms and work with rotations, networkx 2.8.1 \cite{SciPyProceedings_11} for systematic generation of grids in form of networks, and mendeleev 0.9.0 \cite{mendeleev2014} for access to periodic system properties. Additional dependencies used for organisation and display of information are matplotlib 3.3.4 \cite{Hunter:2007}, seaborn 0.11.2 \cite{Waskom2021}, and pandas 1.4.2 \cite{mckinney-proc-scipy-2010}.

\section{Grid algorithms}
\label{app:gridAlgorithms}

All of the following grid algorithms are implemented inside a grid framework that fulfills some general tasks, especially cutting off superfluous points and converting the output into an unified format, described by Unification algorithm (Alg.~\ref{alg:unification}) in combination with an algorithm for ordering grid points (Alg.~\ref{alg:order}).

\begin{algorithm}[H]
\caption{Unification algorithm}
\label{alg:unification}
\begin{algorithmic}
\State \textbf{Input:} integer $N$, function $rotalg$
\State empty list $rotations$
\State find smallest $i \geq N$ that rotalg can generate
\State list rotations $\gets rotalg(i)$
\State convert list $rotations$ to list $points$
\State order points and truncate at $N$
\State \Return list points containing $N$ points on unit sphere
\end{algorithmic}
\end{algorithm}

\begin{algorithm}[H]
\caption{Ordering grid points}\label{alg:order}
\begin{algorithmic}
\State \textbf{Input:} array[$M\times 3$] $grid$, int $N \leq M$
\For{$i$ = 1, 2 \dots $N$}
    \State $selected$ = points in $grid$ up to index $i$
    \State $candidates$ = points in $grid$ not in $selected$
    \For{$c$ in $candidates$}
        \For{$s$ in $selected$}
            \State calculate cosine distance $d = 1-\frac{c \cdot s}{\|c\|\|s\|}$
            \State calculate average cosine distance $\bar{d}$
        \EndFor
    \EndFor
    \State out of $candidates$, select the one with max $\bar{d}$
    \State swap the selected point with the $i^{th}$ element of $grid$
\EndFor
\State \Return $grid$
\end{algorithmic}
\end{algorithm}

Within the unification framework, six algorithms for generation of rotation grid points (Alg.~\ref{alg:seg}-Alg.~\ref{alg:4d_cube}) are available.

\begin{algorithm}[h!]
\caption{Systematic Euler grid}
\label{alg:seg}
\begin{algorithmic}
\State \textbf{Input:} int $m$ \Comment{\textit{ num of points per dimension}}
    \For{$j$ in 1, 2, 3}
        \State create a list of $m$ equally spaced elements between (0, $2\pi$)
    \EndFor
    \State create a $meshgrid$ out of the three lists
\State \Return $meshgrid$ 
\end{algorithmic}
\end{algorithm}

\begin{algorithm}[h!]
\caption{Random Euler angles}\label{alg:rea}
\begin{algorithmic}
\State \textbf{Input:} int $N$ \Comment{\textit{ num of grid points}}
\For{$n$ in 1, 2, \dots $3N$}
    \State create a random number from (0, $2\pi$)
\EndFor
\State reshape random numbers to $N\times 3$ grid
\State \Return $grid$
\end{algorithmic}
\end{algorithm}

\begin{algorithm}[h!]
\caption{Random unit quaternions}\label{alg:ruq}
\begin{algorithmic}
\State \textbf{Input:} int $N$ \Comment{\textit{# num of grid points}}
\For{$n$ in 1, 2, \dots N}
    \For{$i$ in 1, 2, 3}
        \State $r_i$ = random number between (0, 1) 
    \EndFor
    \State $q_1$ = $\sin{(2\pi r_2)} \sqrt{1-r_1}$
    \State $q_2$ = $\cos{(2\pi r_2)} \sqrt{1-r_1}$
    \State $q_3$ = $\sin{(2\pi r_3)} \sqrt{r_1}$
    \State $q_4$ = $\cos{(2\pi r_3)} \sqrt{r_1}$
    \State create quaternion ($q_1$, $q_2$, $q_3$, $q_4$)
\EndFor
\State \Return list of quaternions
\end{algorithmic}
\end{algorithm}

\begin{algorithm}[h!]
\caption{3D cube grid}\label{alg:3d_cube}
\begin{algorithmic}
\State \textbf{Input:} int $M$ \Comment{\textit{# num of division levels}}
\State create a cube centred at origin
\State add vertices of the cube to $grid$
\For{$m$ in 1, 2, \dots $M$}
    \State divide faces on diagonal square lattice
    \State add new points to $grid$
    \State connect new points into new faces
\EndFor
\State normalise $grid$ vectors to length 1
\State \Return $grid$
\end{algorithmic}
\end{algorithm}

\begin{algorithm}[h!]
\caption{Icosahedron grid}\label{alg:ico}
\begin{algorithmic}
\State \textbf{Input:} int $M$ \Comment{\textit{ num of division levels}}
\State create a regular icosahedron centred at origin
\State add vertices of icosahedron to $grid$
\For{$m$ in 1, 2, \dots $M$}
    \State Position a new point at the middle of every edge
    \State Add new points to $grid$
    \State Connect neighbouring new points with new edges
\EndFor
\State normalise $grid$ vectors to length 1
\State \Return $grid$
\end{algorithmic}
\end{algorithm}

A note on the Alg.~\ref{alg:4d_cube}: when systematically sampling quaternions, we avoid double-coverage by selecting only elements with first coordinate $\leq 0$ (half of all generated quaternions). The opposite could also be implemented.

\begin{algorithm}[h!]
\caption{4D cube grid}\label{alg:4d_cube}
\begin{algorithmic}
\State \textbf{Input:} int $m$ \Comment{\textit{ num of points per dimension}}
\For{$i$ in 1, 2, 3, 4}
    \State create a list of $m$ equally spaced elements between (-$\sqrt{\frac{1}{4}}$, $\sqrt{\frac{1}{4}}$)
\EndFor
\State create a $meshgrid$ out of the four lists
\State normalise $meshgrid$ elements to length 1
\State remove $meshgrid$ elements with first coordinate $\leq 0$
\State \Return $meshgrid$
\end{algorithmic}
\end{algorithm}

Grids are saved as sets of 3D vectors to grid points. In Alg.~\ref{alg:grid_rot}, Rodrigues' formula is used to convert grid points to rotation matrices (which can be subsequently transformed to quaternions).
\begin{algorithm}[H]
\caption{From grid points to quaternions}\label{alg:grid_rot}
\begin{algorithmic}
\State \textbf{Input:} $grid$ \Comment{\textit{ set of $N$ points of unit sphere}}
\State $v$ = (0, 0, 1)
\For{$i$ = 1, 2 \dots $N$}
     \State $v'$ = $i$-th element of $grid$
     \State $w$ = $v \times v'$
     \State $s$ = the length of $w$
     \State $c$ = $v \cdot v'$
     \If{$s$ is 0}
         \State $R$=$I_{3\times3}$
     \Else
         \State $R$=$I_{3\times3}$+skew($w$)+skew($w$)$\cdot$skew($w$)$\cdot$(1-$c$)/$s^2$
     \EndIf
     \State transform the rotation matrix $R$ to a quaternion
\EndFor
\State \Return the set of $quaternions$%
\end{algorithmic}
\end{algorithm}






\bibliography{literature}